\documentclass[12pt,leqno]{article}
\usepackage{ectasects}
\usepackage{adjustbox}
\usepackage{bm}
\usepackage{amsfonts}
\usepackage[colorlinks,citecolor=blue,urlcolor=blue,hyperfootnotes=false]{hyperref}
\usepackage{amssymb}
\usepackage[hyperref]{ntheorem}
\usepackage{natbib}
\usepackage{titlesec}
\usepackage{graphicx}
\usepackage{float}
\usepackage[onehalfspacing]{setspace}
\usepackage{theorem}
\usepackage[nomarginpar]{geometry}
\usepackage{booktabs}
\usepackage{threeparttable}
\usepackage{amsmath}
\usepackage{lscape}
\usepackage{scrextend}
\usepackage{relsize}
\usepackage{multicol}
\usepackage{chngcntr}
\usepackage{lipsum}
\usepackage{manyfoot}
\usepackage[sectionbib]{chapterbib}

\providecommand{\U}[1]{\protect\rule{.1in}{.1in}}
\theoremstyle{plain}
\newtheorem{theorem}{Theorem}
\newtheorem{corollary}{Corollary}

\newtheorem{proposition}{Proposition}

\newtheorem{assumption}{Assumption}
\newtheorem{condition}{Condition}
\theorembodyfont{\normalfont}
\newtheorem{Rem}{Remark}

\deffootnote{1em}{1.6em}{\thefootnotemark \enskip}

\newcommand\independent{\protect\mathpalette{\protect\independentT}{\perp}}
\def\independentT#1#2{\mathrel{\rlap{$#1#2$}\mkern2mu{#1#2}}}
\input{tcilatex}
\numberwithin{equation}{section}
\numberwithin{assumption}{section}
\newfootnote{A}

\begin{document}

\begin{center}
{\large {PROGRAM EVALUATION WITH RIGHT-CENSORED DATA} 
\FootnotetextA{}{Date: First version: November 10, 2015. This version: \today. \\ Department of Economics, Vanderbilt University, 415 Calhoun Hall, Nashville, TN, USA. Email:   \href{mailto: pedro.h.santanna@vanderbilt.edu}{pedro.h.santanna@vanderbilt.edu}. I gratefully acknowledge research support from Spanish Plan Nacional de I+D+I, reference
number ECO2014-55858-P.  I thank Sergio Firpo, Brigham Frandsen, Vishal Kamat, Tong Li, Bernhard
Schmidpeter, and seminar participants at the 2016 Latin American 
Workshop in Econometrics of the Econometric Society for helpful comments. R codes for implementing the treatment effects estimators are available from the author. }%
}

{\large \vspace{0.4cm} }

Pedro H. C. Sant'Anna {\normalsize \vspace{0.4cm} }
\end{center}

{\small \noindent {\textsc{Abstract:}}}{\small \ In a unified framework, we
provide estimators and confidence bands for a variety of treatment effects
when the outcome of interest, typically a duration, is subjected to right
censoring. Our methodology accommodates average, distributional, and
quantile treatment effects under different identifying assumptions including
unconfoundedness, local treatment effects, and nonlinear
differences-in-differences. The proposed estimators are easy to implement,
have close-form representation, are fully data-driven upon estimation of
nuisance parameters, and do not rely on parametric distributional
assumptions, shape restrictions, or on restricting the potential treatment
effect heterogeneity across different subpopulations. These treatment
effects results are obtained as a consequence of more general results on
two-step Kaplan-Meier estimators that are of independent interest: we
provide conditions for applying $(i)$ uniform law of large numbers, $(ii)$
functional central limit theorems, and $(iii)$ we prove the validity of the
ordinary nonparametric bootstrap in a two-step estimation procedure where
the outcome of interest may be randomly censored. } \vspace{0.1cm}

\pagenumbering{arabic}


\section{Introduction}

Assessing whether a policy has any effect on a particular outcome has been
one of the main concerns in empirical research. As summarized in \cite%
{Heckman2007} and \cite{Imbens2009}, the focus of the policy evaluation
literature has been mainly confined to situations where the realized outcome
of interest is completely observed for the treated and the control groups.
However, when the outcome variable is subjected to censoring, such inference
procedures may provide misleading conclusions on the effect of the proposed
policy. Important empirical examples of such a setting include the
evaluation of labor market programs on the length of unemployment, of
correctional programs on recidivism of criminal activities, and of clinical
therapy on the survival time.

The main objective and contribution of this paper is to provide a unified
framework to derive estimation and inference procedures for policy
evaluation when the outcome of interest, typically a duration, is subjected
to right-censoring. Our methodology accommodates average, distributional,
and quantile treatment effects in a variety of identifying assumptions such
as selection on observable, cf. \cite{Hirano2003}, \cite{Firpo2007}, and 
\cite{Donald2013}; access to a binary instrumental variable, cf. \cite%
{Imbens1994}, \cite{Abadie2002a}, \cite{Abadie2003} and \cite{Frolich2013};
and access to repeated observations over time, cf. \cite{Athey2006}. To the
best of our knowledge, this paper is the first to propose such broad policy
evaluation tools for right-censored outcomes without relying on parametric
assumptions or shape restrictions.

Our policy evaluation results build on the fact that many treatment effect
measures commonly used can be written as (smooth) functions of moment
equations of the type%
\begin{equation}
\mathbb{E}\left[ \varphi _{z,h_{0}}\left( Y,X,T\right) \right] =\int \varphi
_{z,h_{0}}\left( \bar{y},\bar{x},\bar{t}\right) F\left( d\bar{y},d\bar{x},d%
\bar{t}\right) ,  \label{m-eq}
\end{equation}%
where $Y$ is the outcome of interest, $T$ is the treatment status, and $X$
is a vector of covariates; $\varphi _{z,h_{0}}$ is some integrable function,
potentially indexed by $z$, and by (infinite dimensional) nuisance
parameters $h_{0}$; and $F$ is the joint cumulative distribution function
(CDF). Therefore, our policy evaluation problem can be translated into the
more general task of estimating moments of the type of (\ref{m-eq}).

In the presence of right-censored outcomes, the main challenge in estimating
(\ref{m-eq}) is the fact that $Y$ is not always observed. That is, instead
of observing a random sample $\left\{ Y_{i},X_{i},T_{i}\right\} _{i=1}^{n}$
of $\left( Y,X,T\right) $ as in the \textquotedblleft complete
data\textquotedblright\ setup, one observes $iid$ copies $\left\{
Q_{i},\delta _{i},X_{i},T_{i}\right\} _{i=1}^{n}$ of $(Q,\delta ,X,T)$,
where $Q=\min \left( Y,C\right) $, $\delta =1\left\{ Y\leq C\right\} $, and $%
C$ is a censoring random variable. Right-censoring is a common feature of
duration outcomes, and may arise for different reasons, such as the end of a
follow-up, or drop out. Thus, when estimating (\ref{m-eq}), one must take
into account this data limitation. In fact, ignoring the censoring problem
or restricting the analysis to uncensored observations leads to biased and
inconsistent estimators for (\ref{m-eq}).

To overcome such problems we propose the following two-step procedure. In
the first step, one consistently estimate $h_{0}$ using parametric,
semiparametric or nonparametric methods, and denote such generic estimator
by $\hat{h}_{n}$. In the second step, one plugs $\hat{h}_{n}$ into (\ref%
{m-eq}), and then replace $F$ with $\hat{F}_{n}^{km},$ where $\hat{F}%
_{n}^{km}$ is a nonparametric multivariate extension of the time-honored 
\cite{Kaplan1958} product-limit estimator that naturally address the
censoring issue\footnote{%
Following \cite{VanNoorden2014}, \cite{Kaplan1958} is, based on Thomson
Reuters' Web of Science as 7 October 2014, the most cited paper in
statistics, and the $11^{th}$ most cited paper in all sciences, with 38,600
citations.}. By combining these two steps, we propose to estimate (\ref{m-eq}%
) by%
\begin{equation}
\mathbb{E}_{n}^{km}\left[ \varphi _{z,\hat{h}_{n}}\left( Q,X,T\right) \right]
=\int \varphi _{z,\hat{h}_{n}}\left( \bar{y},\bar{x},\bar{t}\right) \hat{F}%
_{n}^{km}\left( d\bar{y},d\bar{x},d\bar{t}\right) .  \label{2skm-intro}
\end{equation}%
We label the estimator in (\ref{2skm-intro}) as the two-step Kaplan-Meier
(2SKM) estimator.

The 2SKM estimator inherits many attractive features. First, it is very easy
to implement, has a simple close-form representation, is fully data-driven
upon estimation of the nuisance parameters $h_{0}$, and does not depend on
parametric functional form assumptions on the joint distribution $Y$, $X$
and $T$. This last property is in sharp contrast with \cite{Cox1972}
proportional hazard models, or \cite{Buckley1979} accelerated failure time
models, two of the most popular duration models in the literature. Second,
in the absence of censoring, (\ref{2skm-intro}) reduces to the empirical
analogue of (\ref{m-eq}),%
\begin{equation*}
\mathbb{E}_{n}\left[ \varphi _{z,\hat{h}_{n}}\left( Q,X,T\right) \right] =%
\frac{1}{n}\sum_{i=1}^{n}\varphi _{z,\hat{h}_{n}}\left(
Q_{i},X_{i},T_{i}\right) ,
\end{equation*}%
implying that one can interpret our proposal as a natural generalization of
standard two-step estimation procedures such as \cite{Pakes1989} and \cite%
{Chen2003} to situations in which the outcome is censored.

This article contains two sets of new theoretical results on Kaplan-Meier
integrals (\ref{2skm-intro}). First, we present a set of sufficient
conditions under which the 2SKM estimator is uniformly consistent, and
converges weakly to a tight Gaussian process. Furthermore, since the
limiting variance function may depend on the data generating process in
rather complicated forms, we propose and prove the validity of the ordinary
nonparametric bootstrap, which can be used to construct asymptotic valid
confidence bands.

The second set of results deals with estimation and inference under
primitive conditions in three leading policy evaluation methods.
Specifically, we prove that the high-level conditions to establish the
functional central limit theorem and validity of bootstrap hold for average,
distributional, and quantile treatment effects under the unconfoundedness,
local treatment effects, and nonlinear differences-in-differences setups.

This article contributes to the literature on treatment effects with
censored data. Contrary to \cite{Ham1996}, \cite{Eberwein1997}, \cite%
{Hubbard2000}, \cite{Anstrom2001}, \cite{Abbring2003}, and \cite%
{Vanderlaan2003}, our methodology does not rely on parametric models,
separability or proportionality restrictions. In contrast with \cite%
{Frandsen2014}, our proposal can easily accommodate covariates, does not
rely on the potentially restrictive condition that the censoring variable is
always observed, and does not require choosing truncation parameters.
Furthermore, it is important to emphasize that, in contrast to all the
aforementioned proposals, our main results are generic, can be used under a
variety of identification conditions, and apply to any functional of
interest that satisfy the relatively weak conditions.

We also contribute to the literature on Kaplan-Meier integrals, cf. \cite%
{Stute1993a}, \cite{Stute1993, Stute1995a,Stute1996a, Stute1996b, Stute1999}%
, \cite{Wang1999}, \cite{Akritas2000a}, and \cite{Sellero2005}. The
available results in this literature are not directly applicable to our
two-step framework in which the integrand is indexed by unknown, possibly
infinite-dimensional nuisance parameters that have to be estimated
beforehand. Thus, our results for 2SKM estimators complement and extend
those available in the literature.

In order to achieve the aforementioned results, one must bear in mind that
although we do not restrict the dependence between $Y$, $X$ and $T$, our
estimation and inference procedure relies on the maintained assumptions that 
$(a)$ conditionally on the treatment status $T$, the outcome of interest ($Y$%
) is independent of the censoring variable $\left( C\right) $, and $(b)$
conditionally on $Y$ and $T$, the vector of available covariates $(X)$ does
not provide any additional information if censoring will take place. These
assumptions are standard in censoring models, and nest the setups considered
by, e.g. \cite{Powell1986}, \cite{Honore2002}, \cite{Hong2003a}, \cite%
{Lee2005}, \cite{Blundell2007}, and \cite{Frandsen2014}. Nonetheless, these
maintained assumptions are stronger than assuming that, conditionally on $X$
and $T$, $Y$ is independent of $C$, and may be violated in some
applications. Thus, as a form of specification test for 2SKM estimators, it
may be desirable to test our maintained assumptions on the censoring
mechanism. In the supplemental appendix we show that such a task is
feasible, and discuss how one can implement a likelihood ratio type test for
the assumptions. Constructing such a nonparametric test is only feasible at
the cost of introducing additional smoothness and support restrictions on
the underlying data generating process, on top of making use of tuning
parameters such as bandwidths.

The rest of the paper is organized as follows. In Section \ref{motivating}
we motivate the problem at hands by showing that different treatment effects
parameters can be written as smooth functions of moment equations of the
type of (\ref{m-eq}). In Section \ref{intro-km} we discuss the
identification and estimation of generic moments of the type of (\ref{m-eq})
when the outcome of interest is censored. Section \ref{two-step} discusses
some sufficient conditions to derive (uniform) law of large numbers and
(functional) central limit theorems for the proposed 2SKM estimators. We
also discuss some regularity conditions for establishing the validity of the
ordinary nonparametric bootstrap for censored data. In Section \ref{ex} we
use our general results on 2SKM estimators to establish the asymptotic
properties of the treatment effect parameters discussed in Section \ref%
{motivating} in the presence of censored outcomes. In Section \ref{mc} we
conduct a small scale Monte Carlo exercise to illustrate the finite sample
properties of our proposal. Section \ref{conc} concludes with a summary of
the main results. A supplemental appendix includes: $\left( i\right) $ the
proofs of the results herein; $\left( ii\right) $ a discussion on how one
can test the maintained assumptions on the censoring mechanism; and $\left(
iii\right) $ the complete set of Monte Carlo results.

\section{Motivating Examples\label{motivating}}

In this section, we show that, under different identification scenarios, one
can use (smooth) functions of moment equations of the type of (\ref{m-eq})
to characterize the average, distributional, and quantile treatment effects.
We particularly focus on three popular identification setups: $(i)$
selection on observables, $\left( ii\right) $ access to a binary
instrumental variable, and $(iii)$ access to repeated observations over time.

We use the following notation. Let $Y_{0}$ and $Y_{1}$ be the potential
individual outcomes under the control and treatment group, respectively.
Upon inflow, an individual is assigned to a treatment $(T=1)$ or to a
control $\left( T=0\right) $ group. The realized outcome of interest is $%
Y\equiv $ $TY_{1}+(1-T)Y_{0}$, and $X$ is a $k$-dimensional vector of
pre-treatment observable covariates. Let $%
\independent%
$ mean \textquotedblleft is independent\textquotedblright , and $\mathcal{Y}$
denote the support of the random variable $Y$.\bigskip

\textbf{Example 2.1 (}\textit{Unconfoundedness setup). }One of the most
popular identification strategies in policy evaluation is to assume that
selection into treatment is solely based on observable characteristics, i.e. 
$\left( Y_{0},Y_{1}\right) 
\independent%
T|X$~$a.s\mathbf{.}$. This is the so called unconfoundedness setup. Here,
popular parameters of interest are the overall average, distributional, and
quantile treatment effects 
\begin{eqnarray*}
ATE &\equiv &\mathbb{E}\left( Y_{1}-Y_{0}\right) , \\
DTE\left( y\right) &\equiv &\mathbb{P}\left( Y_{1}\leq y\right) -\mathbb{P}%
\left( Y_{0}\leq y\right) , \\
QTE\left( \tau \right) &\equiv &q_{Y_{1}}\left( \tau \right)
-q_{Y_{0}}\left( \tau \right) ,
\end{eqnarray*}%
respectively, where for $t\in \left\{ 0,1\right\} $, $q_{Y_{t}}\left( \tau
\right) \equiv \inf \left\{ y:\mathbb{P}\left( Y_{t}\leq y\right) \geq \tau
\right\} .$

As shown by \cite{Rosenbaum1983}, provided that individuals with the same $X$
values have a positive probability of being both at the treatment and the
control group, the aforementioned treatment effect parameters are identified
by 
\begin{align}
ATE& =\mathbb{E}\left[ \frac{TY}{p\left( X\right) }\right] -\mathbb{E}\left[ 
\frac{\left( 1-T\right) Y}{1-p\left( X\right) }\right] ,  \label{ate} \\
DTE\left( y\right) & =F_{Y_{1}}\left( y\right) -F_{Y_{0}}\left( y\right) ,
\label{dte} \\
QTE\left( \tau \right) & =F_{Y_{1}}^{-1}\left( \tau \right)
-F_{Y_{0}}^{-1}\left( \tau \right) ,  \label{qte}
\end{align}%
where $p\left( X\right) \equiv \mathbb{P}\left( T=1|X\right) $ is the
propensity score, i.e. the probability of selection into treatment,%
\begin{equation*}
F_{Y_{1}}\left( y\right) =\mathbb{E}\left[ \frac{T1\left\{ Y\leq y\right\} }{%
p\left( X\right) }\right] ,~F_{Y_{0}}\left( y\right) =\mathbb{E}\left[ \frac{%
\left( 1-T\right) 1\left\{ Y\leq y\right\} }{1-p\left( X\right) }\right] 
\text{,}
\end{equation*}%
and, for $t\in \left\{ 0,1\right\} $, $F_{Y_{t}}^{-1}\left( \tau \right)
\equiv \inf \left\{ y:F_{Y_{t}}\left( y\right) \geq \tau \right\} $\footnote{%
The average, distributional, and quantile treatment effects on treated
subpopulation can also be identified using a similar strategy.}.

Notice that (\ref{ate}) and (\ref{dte}) are simple differences of moment
equations of the type of (\ref{m-eq}), where, in both cases, $p\left( \cdot
\right) $ plays the role of the unknown nuisance parameter $h_{0}$, and $%
y\in \mathcal{W}\subseteq \mathcal{Y}$ plays the role of $z$ in (\ref{dte}).
Although one cannot write the quantile treatment effects (\ref{qte}) as
moment equations of the type of (\ref{m-eq}), its identification follows
from the one-to-one relationship between the quantile function $%
F_{Y_{t}}^{-1}\left( \tau \right) $ and the CDF $F_{Y_{t}}\left( y\right) $, 
$t\in \left\{ 0,1\right\} $\footnote{%
For estimation and inference purposes, when one is interested in quantile
treatment effects, we will impose additional continuity restrictions on the
DGP, such that the functional delta method can be applied, see e.g. Chapter
3.9 of \cite{VanderVaart1996}. We defer discussion of these assumptions to
Section \ref{ex}.}. Thus, the treatment effect measures (\ref{ate})-(\ref%
{qte}) fit well into our framework.\bigskip

\textbf{Example 2.2 (}\textit{Local treatment effects setup) }In many
circumstances, the assumption that the selection into treatment is based
only on observable characteristics may be unrealistic. \cite{Imbens1994} and 
\cite{Angrist1996} point out that when this is the case and a binary
instrument ($Z)$ for the selection into treatment is available, one can only
nonparametrically identify treatment effect measures for the subpopulation
of compliers, that is, individuals who comply with their actual assignment
of treatment, and would have complied with the alternative assignment. Such
policy evaluation framework is know as the local treatment effect (LTE)
setup.

By following similar arguments as \cite{Rosenbaum1983}, \cite{Abadie2003}
and \cite{Frolich2013} show that, under some regularity conditions to be
discussed in Section \ref{lte-ex}, the average, distributional and quantile
treatment effects for the subpopulation of compliers,%
\begin{align*}
LATE& \equiv \mathbb{E}\left( Y_{1}-Y_{0}|\mathcal{C}\right) , \\
LDTE\left( y\right) & \equiv \mathbb{P}\left( Y_{1}\leq y|\mathcal{C}\right)
-\mathbb{P}\left( Y_{0}\leq y|\mathcal{C}\right) , \\
LQTE\left( \tau \right) & \equiv q_{Y_{1}|\mathcal{C}}\left( \tau \right)
-q_{Y_{0}|\mathcal{C}}\left( \tau \right) ,
\end{align*}%
respectively, can be identified by%
\begin{eqnarray}
LATE &=&\mathbb{E}\left[ Y_{1}^{c}\right] -\mathbb{E}\left[ Y_{0}^{c}\right]
,  \label{late} \\
LDTE\left( y\right) &=&F_{Y_{1}^{c}}\left( y\right) -F_{Y_{0}^{c}}\left(
y\right) ,  \label{ldte} \\
LQTE\left( \tau \right) &=&F_{Y_{1}^{c}}^{-1}\left( \tau \right)
-F_{Y_{0}^{c}}^{-1}\left( \tau \right) ,  \label{lqte}
\end{eqnarray}%
where, for $t\in \left\{ 0,1\right\} $, 
\begin{align}
\mathbb{E}\left[ Y_{t}^{c}\right] & =\frac{1}{\kappa _{t}\left( e\right) }%
\left( \mathbb{E}\left[ \frac{1\left\{ T=t\right\} ZY}{e\left( X\right) }%
\right] -\mathbb{E}\left[ \frac{1\left\{ T=t\right\} \left( 1-Z\right) Y}{%
1-e\left( X\right) }\right] \right) ,  \label{et} \\
F_{Y_{t}^{c}}\left( y\right) & =\frac{1}{\kappa _{t}\left( e\right) }\left( 
\mathbb{E}\left[ \frac{1\left\{ T=t\right\} Z1\left\{ Y\leq y\right\} }{%
e\left( X\right) }\right] -\mathbb{E}\left[ \frac{1\left\{ T=t\right\}
\left( 1-Z\right) 1\left\{ Y\leq y\right\} }{1-e\left( X\right) }\right]
\right) ,  \label{dt}
\end{align}
and%
\begin{equation*}
\kappa _{t}\left( e\right) \equiv \mathbb{E}\left[ \frac{Z1\left\{
T=t\right\} }{e\left( X\right) }-\frac{\left( 1-Z\right) 1\left\{
T=t\right\} }{1\mathbb{-}e\left( X\right) }\right] ,
\end{equation*}
$F_{Y_{t}^{c}}^{-1}\left( \tau \right) =\inf \left\{ y:F_{Y_{t}^{c}}\left(
y\right) \geq \tau \right\} $, and $e\left( X\right) \equiv \mathbb{P}%
(Z=1|X) $.

From (\ref{et}) and (\ref{dt}), one can see that $\mathbb{E}\left[ Y_{t}^{c}%
\right] $ and $F_{Y_{t}^{c}}\left( y\right) $ are scaled differences of
moment equations of the type of (\ref{m-eq}). Analogously to the
unconfoundedness setup, $e\left( \cdot \right) $ plays the role of $h_{0}$,
and $y\in \mathcal{W}\subseteq \mathcal{Y},$ and $\tau \in \left( 0,1\right) 
$ play the role of $z$ in (\ref{ldte}), and (\ref{lqte}), respectively.
Although identification of the aforementioned treatment effects involve $%
\kappa _{t}\left( e\right) $, for estimation and inference purpose, we can
treat $\kappa _{t}\left( e\right) $ as a known function, cf. \cite%
{Abadie2003} and \cite{Frolich2013}. Thus, as in Example 2.1, the treatment
effect measures (\ref{late})-(\ref{lqte}) fit well into our
framework.\bigskip

\textbf{Example 2.3 }\textit{(Differences-in-Differences setup) }This
example is concerned with treatment effects when one has access to repeated
observations over time, the so called differences-in-differences (DID)
approach, cf. \cite{Angrist1999}. In its basic form, a control group is not
treated at two time periods, whereas a treatment group is treated at the
second period. In such a setup, $T=G\cdot I$, $G=\left\{ 0,1\right\} $, $%
I=\left\{ 0,1\right\} ,$ where $G$ is equal to 1 for the treatment group and
0 otherwise, and $I$ is a time indicator such that $I=0$ for the
pre-treatment period and $I=1$ for the post-treatment period. Covariates $X$
are not available.

In this setup, one is usually interested in estimating the average,
distributional, and quantile treatment effects for the treated subpopulation,%
\begin{eqnarray*}
ATT &\equiv &\mathbb{E}\left( Y_{1}-Y_{0}|T=1\right) , \\
DTT\left( y\right)  &\equiv &\mathbb{P}\left( Y_{1}\leq y|T=1\right) -%
\mathbb{P}\left( Y_{0}\leq y|T=1\right) , \\
QTT\left( \tau \right)  &\equiv &q_{Y_{1}|T=1}\left( \tau \right)
-q_{Y_{0}|T=1}\left( \tau \right) .
\end{eqnarray*}

In a seminal work, \cite{Athey2006} show that, although the classical DID
model as in \cite{Card1994} may not be adequate to estimate treatment
effects beyond the average, a generalization of the DID model, the
changes-in-changes (CIC) model, can be used to nonparametrically identify
the $ATT$, $DTT\left( y\right) $ and $QTT\left( \tau \right) $. More
specifically, \cite{Athey2006} show that, under some conditions to be
discussed in Section \ref{CiC-sec},%
\begin{eqnarray}
ATT &=&\mathbb{E}\left[ Y_{11}\right] -\mathbb{E}\left[
F_{Y_{01}}^{-1}(F_{Y_{00}}\left( Y_{10}\right) \right] ,  \label{att} \\
DTT\left( y\right) &=&F_{Y_{11}}\left( y\right) -F_{10}\left(
F_{Y_{00}}^{-1}\left( F_{Y_{01}}\left( y\right) \right) \right) ,
\label{dtt} \\
QTT\left( \tau \right) &=&F_{Y_{11}}^{-1}\left( \tau \right)
-F_{Y_{01}}^{-1}\left( F_{Y_{00}}\left( F_{10}^{-1}(\tau )\right) \right) ,
\label{qtt}
\end{eqnarray}%
where, for $g=\left\{ 0,1\right\} $, $j=\left\{ 0,1\right\} $, $Y_{gj}$ are
the realized outcome $Y$ conditional on $G=g$ and $I=j$, and $%
F_{Y_{gj}}\left( y\right) =\mathbb{E}\left( 1\left\{ Y_{gj}\leq y\right\}
\right) $, and $F_{Y_{gj}}^{-1}\left( \tau \right) =\inf \left\{
y:F_{Y_{gj}}\left( y\right) \geq \tau \right\} $.

Different from previous examples, not all terms in (\ref{att})-(\ref{qtt})
are indexed by unknown functions, and when they do, there is more than one
nuisance function. That is, (\ref{att}) is the difference between $\mathbb{E}%
\left[ Y_{11}\right] ,$ which does not depend on nuisance parameters, and $%
\mathbb{E}\left[ F_{Y_{01}}^{-1}(F_{Y_{00}}\left( Y_{10}\right) \right] $,
where $F_{Y_{01}}^{-1}$ and $F_{Y_{00}}$ play the role of $h$ here. Moving
to (\ref{dtt}), $y$ plays the role of $z$, and $F_{Y_{01}}^{-1}$ and $%
F_{Y_{01}}$ play the role of $h$. Finally, as in Examples 2.1 and 2.2, (\ref%
{qtt}) is a consequence of (\ref{dtt}). Thus, (\ref{att})-(\ref{qtt}) fit
into our framework.

\section{Identification and Estimation of Kaplan-Meier Integrals\label%
{intro-km}}

Let $\left( Y,X,T\right) \in $ $\mathcal{Y\times X\times T\subseteq }$ $%
\mathbb{R}\times \mathbb{R}^{k}\times \left\{ 0,1\right\} $, $F\left(
y,x,t\right) \equiv \mathbb{P}\left( Y\leq y,X\leq x,T\leq t\right) $, and $%
\varphi _{z,h_{0}}\left( Y,X,T\right) $ be a generic known, measurable,
real-valued function indexed by $z\in \mathcal{W\subseteq Y\times X\times T},
$ and by potentially infinite dimensional nuisance parameters $h_{0}\in 
\mathcal{H}$\thinspace , where $\mathcal{H}$ is a Banach space with the
supremum norm. Our goal is to make inference about (\ref{m-eq}), but due to
censoring mechanism, instead of always $Y$, one observes $Q=\min \left(
Y,C\right) $, together with the non-censoring indicator $\delta =1\left\{
Y\leq C\right\} $. Hence, the available data consist of a random sample $%
\left\{ \left( Q_{i},\delta _{i},X_{i},T_{i}\right) \right\} _{i=1}^{n}$
from $\left( Q,\delta ,X,T\right) $, and not $\left\{ \left(
Y_{i},,X_{i},T_{i}\right) \right\} _{i=1}^{n}$ from $\left( Y,X,T\right) $.
In this section, we discuss how one can identify and estimate (\ref{m-eq})
with censored outcomes. Throughout the rest of this paper, all random
variables are defined on a common probability space $\left( \Omega ,\mathcal{%
A},\mathbb{P}\right) $.

We make the following assumption about the censoring mechanism.

\begin{assumption}
\label{censoring_identification}Assume that

\begin{itemize}
\item[(i)] $Y$ $%
\independent%
C|T$

\item[(ii)] $\mathbb{P}\left( \delta =1|X,Y,T\right) =\mathbb{P}\left(
\delta =1|Y,T\right) ~a.s.$.
\end{itemize}
\end{assumption}

Assumption \ref{censoring_identification} states that, conditionally on the
treatment status, the outcome of interest is independent of the censoring
random variable, and that, given the underlying duration\ $Y$ and treatment
status $T$, the covariates do not provide any further information whether
censoring will take place, that is, $\delta $ and $X$ are conditionally
independent given $Y$ and $T$. For instance, a particular case in which
Assumption \ref{censoring_identification} is satisfied is when $C$ is
independent of $\left( Y,X,T\right) $, as assumed by e.g. \cite{Honore2002}, 
\cite{Lee2005}, \cite{Blundell2007}, and \cite{Frandsen2014}. It is
important to have in mind that Assumption \ref{censoring_identification} is
more general than this particular case; it does not impose any restriction
on how $Y$ and $C$ depends on $T$, and it allows some dependency between $C$%
, $T$ and $X.$ Overall, such an assumption is not restrictive when censoring
is fixed, or when the data comes from standard follow-up studies.

Next, we discuss the identification of (\ref{m-eq}) with randomly-censored
data when Assumption \ref{censoring_identification} is satisfied. Denote $%
H_{t}\left( y\right) =\mathbb{P}\left( Q\leq y|T=t\right) $, $G_{t}\left(
y\right) =\mathbb{P}\left( C\leq y|T=t\right) $ and $H_{1t}\left( y,x\right)
=\mathbb{P}\left( Q\leq y,X\leq x,\delta =1|T=t\right) $. Under Assumption %
\ref{censoring_identification}, the joint cumulative hazard function for the
subpopulation with $\left\{ T=t\right\} $ is given by\footnote{%
To see this, note that the probability that a random individual, taken at
random from subpopulation $\left\{ T=t,Y\geq \bar{y}\right\} $, exits the
state of interest\ before $\bar{y}+dy$ and have characteristics $\left\{
X\leq x\right\} $ is $\mathbb{P}\left( \bar{y}\leq Y<\bar{y}+dy,X\leq
x|Y\geq \bar{y},T=t\right) =\left[ F_{t}\left( \left( \bar{y}+dy\right)
-,x\right) -F_{t}\left( \bar{y}-,x\right) \right] /\left[ 1-F_{t}\left( \bar{%
y}-,\infty \right) \right] $. The desired result is achived by integration.}%
\begin{equation*}
\Lambda \left( y,x|T=t\right) =\int_{-\infty }^{y}\frac{F_{t}\left( d\bar{y}%
,x\right) }{1-F_{t}\left( \bar{y}-,\infty \right) },
\end{equation*}%
where $F_{t}\left( y,x\right) \equiv $ $\mathbb{P}\left( Y\leq y,X\leq
x|T=t\right) $ and for any generic function $J$, $J\left( y-\right)
=\lim_{a\uparrow y}J\left( a\right) $, and $J\left\{ y\right\} =J\left(
y\right) -J\left( y-\right) $. For $t\in \left\{ 0,1\right\} $, let $\tau
_{H_{t}}=\inf \left\{ y:H_{t}\left( y\right) =1\right\} $, $\tau
_{F_{t}}=\inf \left\{ y:F_{t}\left( y,\infty \right) =1\right\} $, $\tau
_{G_{t}}=\inf \left\{ y:G_{t}\left( y\right) =1\right\} $ be the least upper
bound of the support of $H_{t}\left( \cdot \right) ,$ $F_{t}\left( \cdot
,\infty \right) $ and $G_{t}\left( \cdot \right) $, respectively. Let $\tau
_{H}=\min \left( \tau _{H_{0}},\tau _{H_{1}}\right) $, and $\tau _{F}$ and $%
\tau _{G}$ are defined analogously.

Next proposition shows that, under Assumption \ref{censoring_identification}%
, we can identify $F\left( y,x,t\right) ,$ which is key to establish the
identification of (\ref{m-eq}). In contrast to \textquotedblleft inverse
probability of censoring\textquotedblright\ (IPC)\ literature, see e.g. \cite%
{Robins1992}, \cite{Vanderlaan2003}, and references therein, our
identification results do not require that $G_{t}\left( \cdot \right) <1$ $%
a.s.,t\in \left\{ 0,1\right\} $, nor relies on continuity assumptions on $Y$
and $C$.

\begin{proposition}
\label{prop-id}Under Assumption \ref{censoring_identification}, for $\left(
y,x,t\right) \in (-\infty ,\tau _{H_{t}}]\times \mathbb{R}^{k}\times \left\{
0,1\right\} ,$ 
\begin{equation*}
\Lambda \left( y,x|T=t\right) =\Lambda ^{cens}\left( y,x|T=t\right) ,
\end{equation*}%
where 
\begin{equation*}
\Lambda ^{cens}\left( y,x|T=t\right) =\int_{-\infty }^{y}\frac{H_{1t}\left( d%
\bar{y},x\right) }{1-H_{t}\left( \bar{y}-\right) }.
\end{equation*}%
Furthermore, for $\left( y,x,t\right) \in (-\infty ,\tau _{H}]\times \mathbb{%
R}^{k}\times \left\{ 0,1\right\} $, 
\begin{equation}
F\left( y,x,t\right) =\sum_{j=0}^{t}\mathbb{P}\left( T=j\right)
\int_{-\infty }^{y}\left( 1-F_{j}\left( \bar{y}-,\infty \right) \right)
\Lambda ^{cens}\left( d\bar{y},x|T=j\right) ,  \label{f-ident}
\end{equation}%
where 
\begin{equation*}
1-F_{t}\left( y,\infty \right) =\exp \left( -\Lambda _{c}^{cens}\left(
y,\infty |T=t\right) \right) \prod\limits_{\bar{y}\leq y}\left( 1-\Lambda
^{cens}\left( \left\{ \bar{y}\right\} ,\infty |T=t\right) \right) ,
\end{equation*}%
and $\Lambda _{c}^{cens}\left( y,x|T=t\right) $ is the continuous part of $%
\Lambda ^{cens}\left( y,x|T=t\right) .$
\end{proposition}

From Proposition \ref{prop-id} one can see that the joint cumulative hazard
plays a major role in the identification of $F\left( y,x,t\right) $. Once we
establish that $\Lambda \left( y,x|T=t\right) $ can be written in terms of $%
\left( Q,\delta ,X,T\right) $, we just need to plug in $\Lambda
^{cens}\left( y,x|T=t\right) $ into (\ref{f-ident}) to recover $F\left(
y,x,t\right) $. Another important implication of Proposition \ref{prop-id}
is that nonparametric identification of $F\left( y,x,t\right) $ over the
entire support of $Y$ may not be feasible. This is intuitive since outcomes
beyond $\tau _{H}=\min \left( \tau _{F},\tau _{G}\right) $ are never
observed for both treatment and control groups. Such restriction is
important, because it implies that the general moment condition (\ref{m-eq})
will be identified only if one of the following conditions holds:

\begin{condition}
\label{cond1}$\tau _{F}\leq \tau _{G}$ where equality may hold except when $%
1-G\left( \tau _{H}-\right) =0$ and $F\left( \left\{ \tau _{H}\right\}
,\infty ,\infty \right) >0$.
\end{condition}

\begin{condition}
\label{cond2}For some $b<\tau _{H}$, $\varphi _{z,h_{0}}\left( Y,X,T\right)
=0$ for $Y\in (b,\tau _{F}].$
\end{condition}

In order to better understand these conditions, notice that Condition \ref%
{cond1} implies that\ $\tau _{H}=\tau _{F}$. It turns out that the support
of the censoring random variable being larger than or equal to the support
of the outcome of interest is a necessary and sufficient condition for
identifying $F\left( y,x,t\right) $ over its entire support. In fact,
Condition \ref{cond1} can only be dispensed for identification of (\ref{m-eq}%
) if $\varphi _{z,h_{0}}$ satisfies Condition \ref{cond2}. When $\tau
_{H}=\tau _{G}<\tau _{F}$ outcomes beyond $\tau _{G}$ are never observed,
and because $\mathbb{P}\left( \tau _{G}<Y\leq \tau _{F}\right) >0$ $a.s.$,
identification of (\ref{m-eq}) can only be attained if $\varphi
_{z,h_{0}}\left( Y,X,T\right) =0$ in $\left[ \tau _{G},\tau _{F}\right] $.
If neither Condition \ref{cond1} nor Condition \ref{cond2} is satisfied, one
can only nonparametrically point-identify a truncated version of (\ref{m-eq}%
). Hence, identification of (\ref{m-eq}) depends mainly on two things: the
support of $Y$ and $C$, and the type of function $\varphi _{z,h_{0}}$ one is
willing to analyze.

Proposition \ref{prop-id} can also be exploited for estimation purposes.
Intuitively, to estimate $F\left( y,x,t\right) $ we need to estimate $%
\Lambda ^{cens}\left( y,x|T=t\right) $ and $\mathbb{P}\left( T=t\right) ,$
and plug in these estimators into (\ref{f-ident}). But notice that $\Lambda
^{cens}\left( y,x|T=t\right) $ only depends on $H_{1t}\left( y,x\right) $
and $H_{t}\left( y\right) $, and both can be estimated by their sample
analogues%
\begin{eqnarray*}
\hat{H}_{n,1t}\left( y,x\right) &=&\frac{1}{n_{t}}\sum_{i=1}^{n}\delta
_{i}1\left\{ Q_{i}\leq y\right\} 1\left\{ X_{i}\leq x\right\} 1\left\{
T_{i}=t\right\} \text{,} \\
\hat{H}_{n,t}\left( y\right) &=&\frac{1}{n_{t}}\sum_{i=1}^{n}1\left\{
Q_{i}\leq y\right\} 1\left\{ T_{i}=t\right\} ,
\end{eqnarray*}%
where, for $t\in \left\{ 0,1\right\} $, $n_{t}=\sum_{i=1}^{n}1\left\{
T_{i}=t\right\} $. Hence, $\Lambda ^{cens}\left( y,x|T=t\right) $ can be
estimated by%
\begin{equation}
\hat{\Lambda}_{n}^{cens}\left( y,x|T=t\right) =\int_{-\infty }^{y}\frac{\hat{%
H}_{n,1t}\left( d\bar{y},x\right) }{1-\hat{H}_{n,t}\left( \bar{y}-\right) }%
=\sum_{i=1}^{n_{t}}\frac{\delta _{\left[ i:n_{t}\right] }1\left\{
Q_{i:n_{t}}\leq y\right\} 1\left\{ X_{\left[ i:n_{t}\right] }\leq x\right\} 
}{n_{t}-i+1},  \label{cmn}
\end{equation}%
where $Q_{1:n_{t}}\leq $ $\cdots \leq Q_{n_{t}:n_{t}}$ are the ordered $Q$%
-values in the subpopulation with $\left\{ T=t\right\} $, and $X_{\left[
i:n_{t}\right] }$, $\delta _{\left[ i:n_{t}\right] }$ are the concomitants
of the $ith$ order statistics in the $t^{th}$ subpopulation, that is, the $X$
and $\delta $ paired with $Q_{i:n_{t}}$. Since $\hat{\Lambda}%
_{n}^{cens}\left( y,x|T=t\right) $ is purely discrete and that $\mathbb{P}%
\left( T=t\right) $ can be estimated by $n_{t}/n$, by plugging (\ref{cmn})
and $n_{t}/n$ into (\ref{f-ident}) we have that 
\begin{equation}
\hat{F}_{n}^{km}\left( y,x,t\right) =\sum_{j=0}^{t}\frac{n_{j}}{n}%
\int_{-\infty }^{y}\prod\limits_{a<\bar{y}}\left( 1-\hat{\Lambda}%
_{n}^{cens}\left( \left\{ a\right\} ,\infty |T=j\right) \right) \hat{\Lambda}%
_{n}^{cens}\left( d\bar{y},x|T=j\right) .  \label{kmhazard}
\end{equation}

Although (\ref{kmhazard}) seems to have a complicated formula, in the next
corollary we show that this is not the case, that (\ref{kmhazard}) can be
written as a simple data-driven weighted average.

\begin{corollary}
\label{prop-idn}The multivariate Kaplan-Meier estimator (\ref{kmhazard}) for 
$F\left( y,x,t\right) $ can be rewritten as%
\begin{equation}
\hat{F}_{n}^{km}\left( y,x,t\right)
=\sum_{j=0}^{t}\sum_{i=1}^{n_{j}}W_{in_{j}}1\left\{ Q_{i:n_{j}}\leq
y\right\} 1\left\{ X_{\left[ i:n_{j}\right] }\leq x\right\} ,  \label{km}
\end{equation}%
where, for $t\in \left\{ 0,1\right\} $, $1\leq i\leq n_{t},$%
\begin{equation*}
W_{in_{t}}=\frac{n_{t}}{n}\frac{\delta _{\lbrack i:n_{t}]}}{n_{t}-i+1}%
\prod_{j=1}^{i-1}\left[ \frac{n_{t}-j}{n_{t}-j+1}\right] ^{\delta _{\left[
j:n_{t}\right] }}.
\end{equation*}
\end{corollary}

Corollary \ref{prop-idn} is important because it shows that, in practice,
one does not need to first estimate $\Lambda ^{cens}\left( y,x|T=t\right) $
to get an estimator for $F\left( y,x,t\right) $. This is automatically
achieved by the weights $W_{in_{t}}$. Additionally, in the absence of
covariates $(x=\infty )$ and treatments ($n_{0}=n$, and $n_{1}=0)$, (\ref%
{kmhazard}) reduces to the time-honored \cite{Kaplan1958} product limit
estimator of $F\left( y,\infty ,\infty \right) ,$ 
\begin{equation*}
\hat{F}_{n}^{km}\left( y\right) =1-\prod\limits_{i=1}^{n}\left( 1-\frac{%
\delta _{\left[ i:n\right] }}{n-i+1}\right) ^{1\left\{ Q_{i:n}\leq y\right\}
},
\end{equation*}%
cf. \cite{Stute1993a} and \cite{Stute1993}. Thus, we argue that (\ref{km})
can be viewed as a multivariate extension of the \cite{Kaplan1958} product
limit estimator, where the treatment status may affect the censoring and the
outcome distribution in an arbitrary way.

With $\hat{F}_{n}^{km}\left( y,x,t\right) $ at hands, one can estimate (\ref%
{m-eq}) by%
\begin{eqnarray}
\mathbb{E}_{n}^{km}\left[ \varphi _{z,\hat{h}_{n}}\left( Q,X,T\right) \right]
&=&\int \varphi _{z,\hat{h}_{n}}\left( \bar{y},\bar{x},\bar{t}\right) \hat{F}%
_{n}^{km}\left( d\bar{y},d\bar{x},d\bar{t}\right)  \notag \\
&=&\sum_{t=0}^{1}\sum_{i=1}^{n_{t}}W_{in_{t}}\varphi _{z,\hat{h}_{n}}\left(
Q_{i:n_{t}},X_{\left[ i:n_{t}\right] },t\right) ,  \label{2skm}
\end{eqnarray}%
where $\hat{h}_{n}$ is a generic first-step estimator for the unknown
nuisance parameter $h_{0}$. The estimator in (\ref{2skm}) is what we refer
as the two-step Kaplan-Meier estimator for (\ref{m-eq}).

It is clear from (\ref{2skm}) that the 2SKM estimator has a close form
representation, does not depend on functional form assumptions on the joint
distribution $Y$, $X$ and $T$, and is fully data-driven upon estimation of
the nuisance parameters $h_{0}$. Furthermore, in the absence of censoring, $%
W_{in_{t}}=n^{-1}$ $a.s.$, implying that (\ref{2skm}) collapses to%
\begin{equation}
\mathbb{E}_{n}\left[ \varphi _{z,\hat{h}_{n}}\left( Q,X,T\right) \right] =%
\frac{1}{n}\sum_{i=1}^{n}\varphi _{z,\hat{h}_{n}}\left(
Q_{i},X_{i},T_{i}\right) ,  \label{m-emp}
\end{equation}%
the sample analogue of (\ref{m-eq}). Hence, one can clearly see that indeed
the 2SKM estimator (\ref{2skm}) is a natural extension of (\ref{m-emp}) to
the cases in which our outcome of interest is subjected to random
right-censoring.

\section{Asymptotic Properties of the Two-Step Kaplan-Meier Estimator\label%
{two-step}}

In this section we derive the asymptotic properties of the 2SKM estimator (%
\ref{2skm}). We adopt the following notation: for a generic set $\mathcal{G}$%
, let $l^{\infty }\left( \mathcal{G}\right) $ be the Banach space of all
uniformly bounded real functions on $\mathcal{G}$ equipped with the uniform
metric $\left\Vert f\right\Vert _{\mathcal{G}}\equiv \sup_{z\in \mathcal{G}%
}\left\vert f\left( z\right) \right\vert $. Let $\mathcal{W}\mathcal{%
\subseteq }$ $\left( -\infty ,\tau _{H}\right) \times \mathbb{R}^{k}$ $%
\times \left\{ 0,1\right\} $. We study the weak convergence of (\ref{2skm})
and related processes as elements of $l^{\infty }\left( \mathcal{W}\right) $%
. Let $\Rightarrow $ denote weak convergence on $\left( l^{\infty }\left( 
\mathcal{W}\right) ,\mathcal{B}_{\infty }\right) $ in the sense of J.
Hoffmann-J$\phi $rgensen, where $\mathcal{B}_{\infty }$ denotes the
corresponding Borel $\sigma $-algebra - cf. \cite{VanderVaart1996}.

For a generic $h\in \mathcal{H},$ $z\in \mathcal{W}$, define%
\begin{equation*}
S^{\varphi }\left( z,h\right) \equiv \mathbb{E}\left[ \varphi _{z,h}\left(
Y,X,T\right) \right] =\int \varphi _{z,h}\left( \bar{y},\bar{x},\bar{t}%
\right) F\left( d\bar{y},d\bar{x},d\bar{t}\right) ,
\end{equation*}%
\begin{equation*}
\hat{S}_{n}^{\varphi }\left( z,h\right) \equiv \mathbb{E}_{n}^{km}\left[
\varphi _{z,h}\left( Y,X,T\right) \right] =\int \varphi _{z,h}\left( \bar{y},%
\bar{x},\bar{t}\right) \hat{F}_{n}^{km}\left( d\bar{y},d\bar{x},d\bar{t}%
\right) .
\end{equation*}%
Therefore, $S^{\varphi }\left( z,h_{0}\right) $ and $\hat{S}_{n}^{\varphi
}\left( z,\hat{h}_{n}\right) $ are respectively equal to the target function
(\ref{m-eq}) and its 2SKM estimator (\ref{2skm}).

In the following, we derive a set of sufficient conditions under which $\hat{%
S}_{n}^{\varphi }\left( z,\hat{h}_{n}\right) $ is uniformly consistent, and
converges weakly to a tight Gaussian process. Furthermore, we show that one
can use the ordinary nonparametric bootstrap to conduct asymptotically valid
inference. These results are novel, complementing and extending those
available in the literature on Kaplan-Meier integrals, cf. \cite{Stute1993a}%
, \cite{Stute1993, Stute1995a, Stute1996a, Stute1996b, Stute2004}, \cite%
{Stute2000}, and \cite{Sellero2005}.

\subsection{Consistency}

For the 2SKM estimator in (\ref{2skm}) to be uniformly consistent, we state
the following sufficient conditions.

\begin{assumption}
\label{cons}$\left\Vert \hat{h}_{n}-h_{0}\right\Vert _{\infty }=o_{\mathbb{P}%
}\left( 1\right) .$
\end{assumption}

\begin{assumption}
\label{continuity}Uniformly in $z\in \mathcal{W},$ $S^{\varphi }\left(
z,h\right) $ is continuous at $h_{0}.$
\end{assumption}

\begin{assumption}
\label{Glivenko} $\left\{ \psi _{1,z,h}:z\in \mathcal{W},h\in H_{\delta
}\right\} $ is Glivenko-Cantelli with an integrable envelope, where $\psi
_{1,z,h}=\varphi _{z,h}\left( Y,X,T\right) $.
\end{assumption}

Assumptions \ref{cons}-\ref{Glivenko} are standard requirements in two-step
estimation procedures, cf. \cite{Chen2003}, and are not related to the
censoring problem. Assumption \ref{cons} requires consistent estimation of
the nuisance parameters $h_{0}$. Assumption \ref{continuity} is a standard
continuity condition, and is weaker than directly imposing a continuity
assumption in $\varphi _{z,h}\left( Y,X,T\right) $. Finally, Assumption \ref%
{Glivenko} put some restrictions on the class of functions $\varphi
_{z,h}\left( Y,X,T\right) .$ Now we state the uniform consistency for $\hat{S%
}_{n}^{\varphi }\left( z,\hat{h}_{n}\right) $.

\begin{theorem}
\label{consistency}Assume that either Condition \ref{cond1} or Condition \ref%
{cond2} holds. Under Assumptions \ref{censoring_identification}, \ref{cons}-%
\ref{Glivenko}, 
\begin{equation*}
\sup_{z\in \mathcal{W}}\left\vert \hat{S}_{n}^{\varphi }\left( z,\hat{h}%
_{n}\right) -S^{\varphi }\left( z,h_{0}\right) \right\vert =o_{\mathbb{P}%
}\left( 1\right) .
\end{equation*}
\end{theorem}

Theorem \ref{consistency} is the first main and new result of the paper. It
shows that under some relatively weak regularity conditions our 2SKM
estimator satisfies a uniform law of large numbers.

\begin{Rem}
\label{rem1}In Theorem \ref{consistency}, Condition \ref{cond1} or Condition %
\ref{cond2} is necessary for identification, but plays no role in deriving
the asymptotic properties of $\hat{S}_{n}^{\varphi }\left( z,\hat{h}%
_{n}\right) $. That is, if neither of these conditions are satisfied, we
have that%
\begin{equation*}
\sup_{z\in \mathcal{W}}\left\vert \int \varphi _{z,\hat{h}_{n}}\left( \bar{y}%
,\bar{x},\bar{t}\right) \hat{F}_{n}^{km}\left( d\bar{y},d\bar{x},d\bar{t}%
\right) -\int \varphi _{z,h_{0}}\left( \bar{y},\bar{x},\bar{t}\right) \check{%
F}\left( d\bar{y},d\bar{x},d\bar{t}\right) \right\vert =o_{\mathbb{P}}\left(
1\right) ,
\end{equation*}%
where 
\begin{equation*}
\check{F}\left( y,x,t\right) =\left\{ 
\begin{array}{cc}
F\left( y,x,t\right) & \text{if }y<\tau _{H} \\ 
F\left( \tau _{H}-,x,t\right) +1\left\{ \tau _{H}\in A\right\} F\left(
\left\{ \tau _{H}\right\} ,x,t\right) & \text{if }y\leq \tau _{H}%
\end{array}%
\right.
\end{equation*}%
and $A$ is the (potentially empty) set of all atoms of $\mathbb{P}\left(
Q\leq \cdot \right) $.
\end{Rem}

\begin{Rem}
\label{rem3} Let $X=\left( X_{disc},X_{cont}\right) ,$ where $X_{disc}$
contains the discrete covariates, and $X_{cont}$ contains the continuous
covariates. At the cost of additional notation, we can relax Assumption \ref%
{censoring_identification} to 
\begin{eqnarray*}
&&\left( i^{\prime }\right) Y%
\independent%
C|T,X_{disc}, \\
&&\left( ii^{\prime }\right) \mathbb{P}\left( \delta =1|X,Y,T\right) =%
\mathbb{P}\left( \delta =1|X_{disc},Y,T\right) ~a.s.,
\end{eqnarray*}%
by standard sample-splitting arguments. Such a strategy can be particularly
useful when the data can be partitioned into groups/clusters (such as
states, cities, schools, etc.), and would not require smoothing parameters,
trimming procedures, and tail behavior restrictions.
\end{Rem}

\subsection{Asymptotic Normality}

To derive the limiting distribution of $\hat{S}_{n}^{\varphi }\left( z,\hat{h%
}_{n}\right) $, we impose the following sufficient conditions:

\begin{assumption}
\label{cons-rate}$h_{n}\in \mathcal{H}$ almost surely and $\left\Vert \hat{h}%
_{n}-h_{0}\right\Vert _{\infty }=o_{\mathbb{P}}\left( n^{-1/4}\right) .$
\end{assumption}

\begin{assumption}
\label{differentiable}The pathwise derivative of $S^{\varphi }\left(
z,h_{0}\right) ,$ $\Gamma ^{\varphi }\left( z,h_{0}\right) [h-h_{0}]$, exist
in all directions $\left[ h-h_{0}\right] \in \mathcal{H}.$ Moreover, for
some constant $0\leq c<\infty ,$%
\begin{equation*}
\left\Vert S^{\varphi }\left( z,h\right) -S^{\varphi }\left( z,h_{0}\right)
-\Gamma ^{\varphi }\left( z,h_{0}\right) [h-h_{0}]\right\Vert _{\infty }\leq
c\left\Vert h-h_{0}\right\Vert _{\infty }^{2}.
\end{equation*}
\end{assumption}

\begin{assumption}
\label{Donsker} The functional class $\left\{ \psi _{1,z,h}:z\in \mathcal{W}%
,h\in H_{\delta }\right\} $ is a Donsker class of functions.
\end{assumption}

\begin{assumption}
\label{delta}Let $\kappa _{z}^{\varphi ,h_{0}}$ be a measurable function
indexed by $z\in \mathcal{W}$ such that $\sup_{z\in \mathcal{W}}\left\vert 
\mathbb{E}\left[ \kappa _{z}^{\varphi ,h_{0}}\left( Q,X,T\right) \right]
\right\vert =0.$ Denote $\kappa _{i}^{\varphi }\left( z,h_{0}\right) \equiv
\kappa _{z}^{\varphi ,h_{0}}\left( Q_{i},X_{i},T_{i}\right) $. Then:

$(i)$ $\sqrt{n}\left( \Gamma ^{\varphi }\left( z,h_{0}\right)
[h_{n}-h_{0}]\right) =n^{-1/2}\sum_{i=1}^{n}\kappa _{i}^{\varphi }\left(
z,h_{0}\right) +o_{\mathbb{P}}\left( 1\right) ,~uniformly$ in $z\in \mathcal{%
W},$

$(ii)\left\{ \psi _{2,z}:z\in \mathcal{W}\right\} $ is a uniformly bounded
Donsker class of functions where $\psi _{2,z}=\kappa _{z}^{\varphi
,h_{0}}\left( Q,X,T\right) .$
\end{assumption}

Assumptions \ref{cons-rate}-\ref{delta} are not related to the censoring
problem, and are standard in two-step estimation procedures, cf. \cite%
{Chen2003}. Assumption \ref{cons-rate} strengthens Assumption \ref{cons}
such that the estimator of the nuisance parameter converges at a rate faster
than $n^{-1/4}$. Assumption \ref{differentiable} is a smooth condition for $%
S^{\varphi }\left( z,h_{0}\right) $ that strengthens Assumption \ref%
{continuity}. Assumption \ref{Donsker} imposes additional restrictions on $%
\varphi _{z,h},$ and it may be verified by using Theorem 3 of \cite{Chen2003}%
, for example. Assumption \ref{delta} is related to the estimation of the
nuisance parameter $h_{0}$, and it is a sufficient condition to $\sqrt{n}%
\left( \Gamma ^{\varphi }\left( z,h_{0}\right) [\hat{h}_{n}-h_{0}]\right) $
converge weakly. It assumes that $\sqrt{n}\left( \Gamma ^{\varphi }\left(
z,h_{0}\right) [h-h_{0}]\right) $ is a smooth linear functional of $%
[h-h_{0}],$ and that one can use a functional central limit theorem in its
linear representation. When $h_{0}$ is consistently estimated by parametric
methods, Assumption \ref{delta} will be satisfied under mild integrability
and smoothness conditions. When $h_{n}$ is nonparametric and has a closed
form expression, under mild conditions, one can use the Riesz representation
approach to obtain $\kappa ^{\varphi }\left( z,h_{0}\right) $. Once $\kappa
^{\varphi }\left( z,h_{0}\right) $ is obtained, Assumption \ref{delta}$(ii)$
can be verified using empirical process theory, cf. \cite{VanderVaart1996}.

It turns out that Assumptions \ref{cons-rate}-\ref{delta} are not sufficient
to derive the asymptotic distribution of $\hat{S}_{n}^{\varphi }\left( z,%
\hat{h}_{n}\right) .$We need some additional conditions due to the censoring
problem. Define 
\begin{multline}
\eta _{i}^{\varphi }\left( z,h_{0}\right) =\sum_{t=0}^{1}1\left\{
T_{i}=t\right\} [\varphi _{z,h_{0}}\left( Q_{i},X_{i},T_{i}\right) \gamma
_{0t}\left( Q_{i}\right) \delta _{i}  \label{rep} \\
+\gamma _{1t,z,h_{0}}^{\varphi }\left( Q_{i}\right) \left( 1-\delta
_{i}\right) -\gamma _{2t,z,h_{0}}^{\varphi }\left( Q_{i}\right) ],
\end{multline}%
where, for a generic $h\in \mathcal{H}$, 
\begin{eqnarray*}
\gamma _{0t}\left( y\right) &=&\exp \left\{ \int_{0}^{y-}\frac{H_{0t}\left( d%
\bar{w}\right) }{1-H_{t}\left( \bar{w}\right) }\right\} , \\
\gamma _{1t,z,h}^{\varphi }\left( y\right) &=&\frac{1}{1-H_{t}\left(
y\right) }\int 1\left\{ y<\bar{w}\right\} \varphi _{z,h}\left( \bar{w},\bar{x%
},t\right) \gamma _{0t}\left( \bar{w}\right) H_{t}\left( d\bar{w},d\bar{x}%
\right) , \\
\gamma _{2t,z,h}^{\varphi }\left( y\right) &=&\int \int \frac{1\left\{ \bar{v%
}<y,\bar{v}<\bar{w}\right\} \varphi _{z,h}\left( \bar{w},\bar{x},t\right) }{%
\left[ 1-H_{t}\left( \bar{v}\right) \right] ^{2}}\gamma _{0t}\left( \bar{w}%
\right) H_{0t}\left( d\bar{v}\right) H_{1t}\left( d\bar{w},d\bar{x}\right) ,
\end{eqnarray*}%
and $H_{t}\left( y\right) $ and $H_{1t}\left( y,x\right) $ are defined as
before and $H_{0t}\left( y\right) =\mathbb{P}\left( Q\leq y,\delta
=0|T=t\right) .$

\begin{assumption}
\label{integrability2}For $t\in \left\{ 0,1\right\} $, $\sup_{z\in \mathcal{W%
}}\left\vert \int \left( \varphi _{z,h_{0}}\left( \bar{y},\bar{x},t\right)
\gamma _{0t}\left( \bar{y}\right) \right) ^{2}H_{1t}\left( d\bar{y},d\bar{x}%
\right) \right\vert <\infty .$
\end{assumption}

\begin{assumption}
\label{bias}For $t\in \left\{ 0,1\right\} $, $\sup_{z\in \mathcal{W}%
}\left\vert \int \left\vert \varphi _{z,h_{0}}\left( \bar{y},\bar{x}%
,t\right) \right\vert C_{t}^{1/2}\left( \bar{y}\right) F_{t}\left( d\bar{y},d%
\bar{x}\right) \right\vert <\infty ,$ $where$%
\begin{equation*}
C_{t}\left( y\right) =\int_{-\infty }^{y-}\frac{G_{t}\left( d\bar{y}\right) 
}{\left[ 1-H_{t}\left( \bar{y}\right) \right] \left[ 1-G_{t}\left( \bar{y}%
\right) \right] }.
\end{equation*}
\end{assumption}

Assumption \ref{integrability2} is a modified \textquotedblleft finite
second moment\textquotedblright\ condition for censored data. In the absence
of censoring, such condition reduces to $\sup_{z}\left\vert \mathbb{E}\left[
\varphi _{z,h_{0}}\left( Y,X,T\right) ^{2}\right] \right\vert <\infty $.
Assumption \ref{integrability2} guarantees that (\ref{rep}) has a finite
variance. Assumption \ref{bias} is to control the bias of the $\hat{S}%
_{n}^{\varphi }\left( z,h_{0}\right) $. Although the bias of $\hat{S}%
_{n}^{\varphi }\left( z,h_{0}\right) $ converges to $0$, the rate of
convergence may be faster than $\sqrt{n}$, and Assumption \ref{bias}
guarantees that the bias is of the order $o\left( n^{-1/2}\right) $. This
issue has been discussed in detail in \cite{Stute1994a}. Whenever Condition %
\ref{cond2} is satisfied, Assumptions \ref{integrability2} and \ref{bias}
will be satisfied provided that $\sup_{z}\left\vert \mathbb{E}\left[ \varphi
_{z,h_{0}}\left( Y,X,T\right) ^{2}\right] \right\vert <\infty $, which is
implied by Assumption \ref{Donsker}. However, this is not necessarily the
case for a generic $\varphi _{z,h_{0}}$ when Condition \ref{cond2} is not
satisfied.

Next theorem presents the weak convergence result for the 2SKM estimator.

\begin{theorem}
\label{weak}Assume that either Condition \ref{cond1} or Condition \ref{cond2}
holds. Suppose that $\hat{S}_{n}^{\varphi }\left( z,\hat{h}_{n}\right) $ is
a uniformly consistent estimator of $S^{\varphi }\left( z,h_{0}\right) $,
and that Assumptions \ref{censoring_identification}, \ref{cons-rate}-\ref%
{bias} are satisfied. Then, in $l^{\infty }\left( \mathcal{W}\right) $, 
\begin{equation*}
\sqrt{n}\left( \hat{S}_{n}^{\varphi }\left( z,\hat{h}_{n}\right) -S^{\varphi
}\left( z,h_{0}\right) \right) \Rightarrow \mathbb{G},
\end{equation*}%
where $\mathbb{G}$ is a tight Gaussian process with mean zero and covariance
function%
\begin{equation}
V^{\varphi }\left( z_{1},z_{2}\right) =\mathbb{E}\left[ \Psi ^{\varphi
}\left( z_{1}\right) \Psi ^{\varphi }\left( z_{2}\right) ^{\prime }\right] ,
\label{var}
\end{equation}%
with 
\begin{equation*}
\Psi ^{\varphi }\left( z\right) \equiv \eta ^{\varphi }\left( z,h_{0}\right)
-S^{\varphi }\left( z,h_{0}\right) +\kappa ^{\varphi }\left( z,h_{0}\right) .
\end{equation*}
\end{theorem}

\begin{Rem}
\label{rem2}Again, Condition \ref{cond1} or Condition \ref{cond2} is only
necessary for identification. If neither is satisfied, one can still derive
the weak convergence of%
\begin{equation*}
\sqrt{n}\left( \int \varphi _{z,\hat{h}_{n}}d\hat{F}_{n}^{km}-\int \varphi
_{z,h_{0}}d\check{F}\right) ,
\end{equation*}%
where $\check{F}$ is defined as in Remark \ref{rem1}. Furthermore, the
arguments in Remark \ref{rem3} also apply here.
\end{Rem}

Theorem \ref{weak} is the second main and new result of the paper. It shows
that under some relatively weak regularity conditions our 2SKM estimator
satisfies a functional central limit theorem. This result forms the basis of
all inference results on policy evaluation with censored data.

As an application of the result above, we can show that plug-in estimators
of Hadamard differentiable functionals also satisfy functional central limit
theorems. Examples include quantile curves, as well as Lorenz curves, and
Gini coefficients.

\begin{corollary}
\label{delta-method}Consider the parameter $\theta $ as an element of a
parameter space $\mathbb{W}_{\theta }\subset l^{\infty }\left( \mathcal{W}%
\right) $, with $\mathbb{W}_{\theta }$ containing the true value of $\theta
_{0}=S^{\varphi }\left( z,h_{0}\right) $. Consider the plug-in estimator $%
\hat{\theta}=\hat{S}_{n}^{\varphi }\left( z,\hat{h}_{n}\right) $ as defined
in (\ref{2skm}). Suppose a functional of interest $\phi :\mathbb{W}_{\theta
}\mapsto l^{\infty }\left( \mathcal{S}\right) $ is Hadamard differentiable
in $\theta $ at $\theta _{0}$ tangentially to $\mathcal{C}\left( \mathcal{W}%
\right) $ with derivative $\phi _{\theta }^{\prime }.$ Then, under the
conditions of Theorem \ref{weak},%
\begin{equation*}
\sqrt{n}\left( \phi \left( \hat{\theta}\right) -\phi \left( \theta \right)
\right) \Rightarrow \phi _{\theta }^{\prime }\left( \mathbb{G}\right) .
\end{equation*}
\end{corollary}

\subsection{Bootstrap\label{sec-boots}}

From Theorem \ref{weak} we have that the asymptotic covariance function (\ref%
{var}) depends on the underlying data generating process and standardization
can be complicated. To see this, note that in order to estimate $V^{\varphi
}\left( \cdot ,\cdot \right) $, one needs to estimate $\eta ^{\varphi }$ and 
$\kappa ^{\varphi }$, plug in our estimator for $S^{\varphi }\left(
z,h_{0}\right) $, $\hat{S}_{n}^{\varphi }\left( z,\hat{h}_{n}\right) ,$ and
then compute the sample second moment of these quantities. But in order to
estimate $\eta ^{\varphi }$ one needs to estimate $\gamma _{0t}$, $\gamma
_{1t,z,h}^{\varphi }$ and $\gamma _{2t,z,h}^{\varphi },t\in \left\{
0,1\right\} $. Furthermore, different estimators of $\kappa ^{\varphi }$
could be needed depending on how one chooses to estimate $h_{0}$. It turns
out that estimating these nuisance functions can be difficult, and may
involve tuning parameters such as bandwidths, cf. \cite{SantAnna2016a}. To
avoid these issues, we follow an alternative route and use the ordinary
nonparametric bootstrap to conduct asymptotically valid inference.

In order to compute the bootstrap confidence bands, let $B$ be a large
integer. For each $b=1,\dots ,B:$

\begin{enumerate}
\item Obtain the resample data $\left\{ Q_{i}^{b},X_{i}^{b},T_{i}^{b},\delta
_{i}^{b}\right\} _{i=1}^{n}.$

\item Estimate 
\begin{equation*}
\hat{S}_{n}^{b,\varphi }\left( z,\hat{h}_{n}^{b}\right)
=\sum_{t=0}^{1}\sum_{i=1}^{n_{t}}W_{in_{t}}^{b}\varphi _{z,\hat{h}%
_{n}^{b}}\left( Q_{i:n_{t}}^{b},X_{\left[ i:n_{t}\right] }^{b},t\right) ,
\end{equation*}%
where $\hat{h}_{n}^{b}$ and $W_{in_{t}}^{b}$ are the same as $\hat{h}_{n}$
and $W_{in_{t}}$ but based on the bootstrap data.

\item Set $L^{b,\varphi }=\max_{z\in \mathcal{W}}\left\vert \hat{S}%
_{n}^{b,\varphi }\left( z,\hat{h}_{n}^{b}\right) -\hat{S}_{n}^{\varphi
}\left( z,\hat{h}_{n}\right) \right\vert $
\end{enumerate}

Then, the $\left( 1-\alpha \right) 100\%$ asymptotic confidence band is
calculated as 
\begin{equation*}
\hat{S}_{n}^{\varphi }\left( z,\hat{h}_{n}\right) -c_{1-\alpha
}^{B}n^{-1/2}\leq S^{\varphi }\left( z,h_{0}\right) \leq \hat{S}%
_{n}^{\varphi }\left( z,\hat{h}_{n}\right) +c_{1-\alpha }^{B}n^{-1/2},
\end{equation*}%
where $c_{1-\alpha }^{B}$ denotes the empirical $\left( 1-\alpha \right) $
quantile of the simulated sample $\left\{ L^{b,\varphi }\right\} _{b=1}^{B}$%
. In practice, the maximum in step 3 is taken over a discretized subset $%
\mathcal{W}.$

Next, we establish the asymptotic validity of the aforementioned bootstrap
procedure considering the following additional conditions on the nuisance
parameters. Here and subsequently, superscript $\ast $ denotes probability
or moment computed under the bootstrap distribution conditional on the
original data set.

\begin{assumption}
\label{boot1}With $\mathbb{P}^{\ast }-probability$ tending to one, $\hat{h}%
_{n}^{\ast }\in \mathcal{H}$, and $\left\Vert \hat{h}_{n}^{\ast }-\hat{h}%
_{n}\right\Vert _{\infty }=o_{\mathbb{P}^{\ast }}\left( n^{-1/4}\right) .$
\end{assumption}

\begin{assumption}
\label{boot3}$\sqrt{n}\left( \Gamma ^{\varphi }\left( z,\hat{h}_{n}\right) [%
\hat{h}_{n}^{\ast }-\hat{h}_{n}]\right) =n^{-1/2}\sum_{i=1}^{n}\kappa
_{i}^{\ast ,\varphi }\left( z,\hat{h}_{n}\right) +o_{\mathbb{P}^{\ast
}}\left( 1\right) ,$ where $\kappa _{i}^{\ast ,\varphi }\left( z,\hat{h}%
_{n}\right) \equiv \kappa _{z}^{\varphi ,\hat{h}_{n}}\left( Q_{i}^{\ast
},X_{i}^{\ast },T_{i}^{\ast }\right) ,$ and $\sup_{z\in \mathcal{W}%
}\left\vert \mathbb{E}^{\ast }\left[ \kappa _{z}^{\varphi ,\hat{h}%
_{n}}\left( Q^{\ast },X^{\ast },T^{\ast }\right) \right] \right\vert =0.$
Furthermore, $\left\{ \kappa ^{\varphi }\left( z,h\right) :z\in \mathcal{W}%
\text{, }h\in \mathcal{H}_{\delta _{n}}\right\} $ is a uniformly bounded
Donsker class of functions.
\end{assumption}

\begin{theorem}
\label{boot}Assume that Assumptions \ref{censoring_identification}, \ref%
{Donsker}, \ref{integrability2} and \ref{bias} hold. Let Assumption \ref%
{differentiable} be satisfied with $h_{0}$ replaced by $h\in \mathcal{H}%
_{\delta _{n}}$. Then, under Assumptions \ref{boot1} and \ref{boot3}, 
\begin{equation*}
\sqrt{n}\left( \hat{S}_{n}^{\ast ,\varphi }\left( z,\hat{h}_{n}^{\ast
}\right) -\hat{S}_{n}^{\varphi }\left( z,\hat{h}_{n}\right) \right) \underset%
{\ast }{\Rightarrow }\mathbb{G},
\end{equation*}%
where $\mathbb{G}$ is the same Gaussian process of Theorem \ref{weak} and $%
\underset{\ast }{\Rightarrow }$ denoting weak convergence in probability
under the bootstrap law - see \cite{Gine1990}.
\end{theorem}

Theorem \ref{boot} is the third main and new result of the paper. It shows
that the limiting distribution of the bootstrap estimator is the same as
that of Theorem \ref{weak}, and hence, our proposed resample scheme is able
to mimic the asymptotic distribution of interest. Such a result is very
powerful and allows one to use the ordinary nonparametric bootstrap to
conduct asymptotically valid inference.

By combining Theorem \ref{boot} with the functional delta method for the
bootstrap, cf. Theorem 3.9.11 in \cite{VanderVaart1996}, we can show the
bootstrap validity of plug-in estimators of Hadamard differentiable
functionals as well.

\begin{corollary}
\label{delta-method-boot}Consider the parameter $\theta $ as an element of a
parameter space $\mathbb{W}_{\theta }\subset l^{\infty }\left( \mathcal{W}%
\right) $, with $\mathbb{W}_{\theta }$ containing the true value of $\theta
_{0}=S^{\varphi }\left( z,h_{0}\right) $. Consider the plug-in estimator $%
\hat{\theta}=\hat{S}_{n}^{\varphi }\left( z,\hat{h}_{n}\right) $ as defined
in (\ref{2skm}), and its bootstrap analogue $\hat{\theta}^{\ast }=\hat{S}%
_{n}^{\ast ,\varphi }\left( z,\hat{h}_{n}^{\ast }\right) .$ Suppose a
functional of interest $\phi :\mathbb{W}_{\theta }\mapsto l^{\infty }\left( 
\mathcal{S}\right) $ is Hadamard differentiable in $\theta $ at $\theta _{0}$
tangentially to $\mathcal{C}\left( \mathcal{W}\right) $ with derivative $%
\phi _{\theta }^{\prime }.$ Then, under the conditions of Theorem \ref{boot},%
\begin{equation*}
\sqrt{n}\left( \phi \left( \hat{\theta}^{\ast }\right) -\phi \left( \hat{%
\theta}\right) \right) \underset{\ast }{\Rightarrow }\phi _{\theta }^{\prime
}\left( \mathbb{G}\right) .
\end{equation*}
\end{corollary}

\section{Treatment Effects with Right-Censored Data\label{ex}}

In this section we illustrate the general applicability of our 2SKM approach
by revisiting the motivating examples of Section \ref{motivating}. In short,
we show that, under relatively weak regularity conditions, one can
consistently estimate, and construct asymptotically valid confidence bands
for the average, distributional, and quantile treatment effects discussed in
Examples 2.1, 2.2 and 2.3 when the outcomes is randomly censored. These
results are novel to the literature, and are obtained by verifying the
high-level conditions in Theorems \ref{consistency}-\ref{boot}.

We use the same potential outcome notation as in Section \ref{motivating},
but due to the censoring mechanisms, instead of observing $Y$, one observes $%
Q\equiv TQ_{1}+(1-T)Q_{0},$ where $Q_{0}=\min \left\{ Y_{0},C_{0}\right\} $, 
$Q_{1}=\min \left\{ Y_{1},C_{1}\right\} $, $C_{0}$ and $C_{1}$ being
potential censoring random variables under the control and treatment groups,
respectively. In addition to $Q$, one also observes the censoring indicator $%
\delta \equiv T\delta _{1}+\left( 1-T\right) \delta _{0}$, where, for $t\in
\left\{ 0,1\right\} $, $\delta _{t}=1\left\{ Y_{t}\leq C_{t}\right\} $. It
is important to emphasize that, in the following, we can accommodate
covariates, allow the treatment status to affect the censoring variable in
an arbitrary way, and we do not impose the potentially restrictive condition
that censoring variable $C$ is always observed.

\subsection{Unconfoundedness Setup\label{unc-ex}}

We first revisit unconfoundedness setup discussed in Example 2.1. We impose
the following conditions.

\begin{assumption}
\label{unconfound}Assume $(i)\left( Y_{0},Y_{1},C_{0},C_{1}\right) 
\independent%
T|X$~$a.s\mathbf{.;}$ $(ii)$ for some $\varepsilon >0$, $\varepsilon
<p\left( \cdot \right) <1-\varepsilon ~a.s.,$ where $p\left( X\right) \equiv 
\mathbb{P}\left( T=1|X\right) $; $(iii)$ $\left( Y_{0},Y_{1}\right) $ $%
\independent%
\left( C_{0},C_{1}\right) |T$; $\left( iv\right) for~t\in \left\{
0,1\right\}, \mathbb{P}\left( \delta _{t}=1|X,T,Y_{t}\right) =\mathbb{P}%
\left( \delta _{t}=1|Y_{t},T\right) ~a.s.$.
\end{assumption}

Assumptions \ref{unconfound}$(i)$ and $(ii)$ are standard in the literature,
cf. \cite{Rosenbaum1983}, \cite{Hirano2003}, \cite{Ichimura2005}, \cite%
{Firpo2007}, \cite{Donald2013}, among others. If censoring is not present,
Assumptions \ref{unconfound}$(i)$ and $(ii)$ suffice to identify our
treatment effects of interest. Nonetheless, censoring introduces another
source of confounding because the probability of censoring is related to
potential outcomes. This additional identification challenge can be overcome
under Assumption \ref{unconfound}$(iii)$, the analogous of Assumptions \ref%
{censoring_identification} in the unconfoundedness context\footnote{%
As discussed in Supplemental Appendix, such an assumption is
testable as long as one imposes additional smoothness and support
restrictions in the DGP.}.

In the absence of censoring, \cite{Rosenbaum1983}, \cite{Hirano2003}, \cite%
{Ichimura2005}, \cite{Firpo2007}, \cite{Donald2013}, among others, have
proposed estimators for (\ref{ate})-(\ref{qte}), where one first estimate $%
p\left( \cdot \right) $ by parametric or nonparametric methods, plugs it
into (\ref{ate})-(\ref{qte}), and then use the analogy principle to estimate
(\ref{ate})-(\ref{qte}). As we have seen in Section \ref{intro-km}, although
such a procedure is not feasible when $Y$ is subject to censoring
mechanisms, one can use the 2SKM procedure to overcome this issue. That is,
under Assumption \ref{unconfound}, one can use the 2SKM methodology, and
estimate (\ref{ate})-(\ref{qte}) by%
\begin{align}
ATE_{n}^{km}& =\mathbb{E}_{n}^{km}\left( Y_{1}\right) -\mathbb{E}%
_{n}^{km}\left( Y_{0}\right) ,  \label{atekm} \\
DTE_{n}^{km}\left( y\right) & =\hat{F}_{n,Y_{1}}^{km}\left( y\right) -\hat{F}%
_{n,Y_{0}}^{km}\left( y\right) ,  \label{dtekm} \\
QTE_{n}^{km}\left( \tau \right) & =\hat{F}_{n,Y_{1}}^{km,-1}\left( \tau
\right) -\hat{F}_{n,Y_{0}}^{km,-1}\left( \tau \right) ,  \label{qtekm}
\end{align}%
respectively, where%
\begin{eqnarray*}
\mathbb{E}_{n}^{km}\left( Y_{1}\right) &=&\sum_{i=1}^{n_{1}}W_{in_{1}}\frac{%
Q_{i:n_{1}}}{\hat{p}_{n}\left( X_{\left[ i:n_{1}\right] }\right) },\text{ }%
\mathbb{E}_{n}^{km}\left( Y_{0}\right) =\sum_{i=1}^{n_{0}}W_{in_{0}}\frac{%
Q_{i:n_{0}}}{1-\hat{p}_{n}\left( X_{\left[ i:n_{0}\right] }\right) }, \\
\hat{F}_{n,Y_{1}}^{km}\left( y\right) &=&\sum_{i=1}^{n_{1}}W_{in_{1}}\frac{%
1\left\{ Q_{i:n_{1}}\leq y\right\} }{\hat{p}_{n}\left( X_{\left[ i:n_{1}%
\right] }\right) },\text{ }\hat{F}_{n,Y_{0}}^{km}\left( y\right)
=\sum_{i=1}^{n_{0}}W_{in_{0}}\frac{1\left\{ Q_{i:n_{0}}\leq y\right\} }{1-%
\hat{p}_{n}\left( X_{\left[ i:n_{0}\right] }\right) },
\end{eqnarray*}%
$\hat{F}_{n,Y_{t}}^{km,-1}\left( \tau \right) $ is the empirical $\tau $%
-quantile of the rearrangement of $\hat{F}_{n,Y_{t}}^{km}\left( y\right) $
if $\hat{F}_{n,Y_{t}}^{km}\left( y\right) $ is not monotone, cf. \cite%
{Chernozhukov2010}, $t\in \left\{ 0,1\right\} $, and $\hat{p}_{n}\left(
\cdot \right) $ is a first-step estimator for the propensity score $p\left(
\cdot \right) .$ Here, for $1\leq i\leq n_{1}$, $Q_{i:n_{1}}$ is the $ith$
order statistics in the treated subsample, and $X_{\left[ i:n_{1}\right] }$
is the concomitants of the $ith$ order statistics in the treated
subpopulation; $Q_{i:n_{0}}$ and $X_{\left[ i:n_{0}\right] }$ are defined
analogously but for the control subsample.

In practice, one can estimate $p\left( \cdot \right) $ by parametric,
semi-parametric or nonparametric methods, e.g. \cite{Rosenbaum1983}, \cite%
{Hahn1998}, \cite{Hirano2003} and \cite{Ichimura2005}. Nonetheless, it is
important to have in mind that different regularity conditions might be
needed depending on the estimation method you use. In the Appendix we
discuss these conditions for three popular estimators of $p\left( \cdot
\right) $: the parametric estimator (e.g. Logit or Probit specifications),
the nonparametric leave-one-out Nadaraya-Watson kernel-based estimator, and
the nonparametric Logit Series estimator. We can show that as long as the
required regularity (smooth) conditions are met, the 2SKM estimators (\ref%
{atekm})-(\ref{qtekm}) are uniform consistent, converge weakly, and the
ordinary nonparametric bootstrap procedure can be used to conduct
asymptotically valid inference. These results are summarized in the next
proposition.

\begin{proposition}
\label{col1}Let Assumption \ref{unconfound} be satisfied. Assume that either
Assumption \ref{parametric}, Assumption \ref{kernel}, or Assumption \ref%
{series} in the Appendix \ref{pscore-cond} is satisfied. Then:

$(i)$ If $Y_{1}$ and $Y_{0}$ are square integrable, Condition \ref{cond1}
holds, and Assumptions \ref{integrability2}-\ref{bias} are satisfied for $%
\varphi _{z,h_{0}}\left( \cdot ,\cdot ,\cdot \right) =ytp\left( x\right)
^{-1}-y\left( 1-t\right) \left( 1-p\left( x\right) \right) ^{-1}$, the $ATE$
estimator (\ref{atekm}) for (\ref{ate}) is $\sqrt{n}$-consistent and
asymptotically normal, and the bootstrap procedure described in Theorem \ref%
{boot} is asymptotically valid.

$(ii)$ The $DTE\left( \cdot \right) $ estimator (\ref{dtekm}) for (\ref{dte}%
) converges weakly in $l^{\infty }\left( \mathcal{Y}\cap (-\infty ,\tau
_{H})\right) ,$ and the bootstrap procedure described in Theorem \ref{boot}
is asymptotically valid.

$(iii)$ Let $\bar{\tau}=\min_{t}\left\{ F_{t}\left( \tau _{H},\infty \right)
\right\} $. If $Y_{0}$ and $Y_{1}$ are absolutely continuous random
variables such that, for $t\in \left\{ 0,1\right\} ,$ $0<\inf_{\tau \in
\left( 0,\bar{\tau}\right) }f_{Y_{t}}\left( F_{Y_{t}}^{-1}\left( \tau
\right) \right) <\sup_{\tau \in \left( 0,\bar{\tau}\right) }f_{Y_{t}}\left(
F_{Y_{t}}^{-1}\left( \tau \right) \right) <\infty $, the $QTE\left( \cdot
\right) $ estimator (\ref{qtekm}) for (\ref{qte}) converges weakly in $%
l^{\infty }(\left( 0,\bar{\tau})\right) $, and the bootstrap procedure
described in Theorem \ref{boot} is asymptotically valid.
\end{proposition}

The results in Proposition \ref{col1} are new to the literature. To the best
of our knowledge, the only available results related to Proposition \ref%
{col1} are \cite{Hubbard2000}, who, for a fixed $y$, proposes an alternative
estimator for the $DTE\left( y\right) $ that relies on a parametric
specification for the propensity score, and \cite{Anstrom2001} who builds on 
\cite{Hubbard2000} and proposes an estimator for the $ATE$. Nonetheless, it
is important to notice that the results in Proposition \ref{col1} go beyond
this particular case: it allows one to use nonparametric estimators of the
propensity score, and justify the use of the bootstrap to conduct uniform
asymptotically valid inference. On one hand, allowing for the propensity
score to be estimated by nonparametric methods can be particularly important
for two reasons: $(a)$ as shown by \cite{Huber2013}, misspecification of the
propensity score may lead to severe distortion on the policy evaluation
parameters of interest; and $(b)$ as shown by \cite{Hirano2003} and \cite%
{Chen2008}, even when the propensity score is correctly specified, using
nonparametric estimates can lead to efficiency gains. On the other hand,
since our bootstrapped confidence sets are uniformly valid in the sense that
they cover the entire functional of interest with pre-specified probability,
they can be used to test functional hypotheses such as no-effect, positive
effect, or stochastic dominance, cf. \cite{Abadie2002}.

\subsection{Local Treatment Effects Setup\label{lte-ex}}

This section proposes and derives the asymptotic properties of 2SKM
estimators of the local average, distributional and quantile treatment
effects described in Example 2.2. To do so, we need to introduce additional
notation. Let $Y_{0},Y_{1},Q_{1},Q_{0},\delta _{1},\,\delta _{0},C_{0},C_{1}$
and $X$ be defined as in the unconfoundedness framework. The local treatment
effect (LTE) setup presumes the availability of a binary instrumental
variable $Z$ for the treatment assignment. Denote $T_{0}$ and $T_{1}$ the
values that $T$ would have taken if $Z$ is equal to zero or one,
respectively. The realized treatment is $T=ZT_{1}+\left( 1-Z\right) T_{0}.$
Thus, the observed sample consist of $iid$ copies $\left\{ Q_{i},\delta
_{i},X_{i},T_{i},Z_{i}\right\} _{i=1}^{n}$ of $(Q,\delta ,X,T,Z)$. Denote $%
e\left( X\right) \equiv \mathbb{P}(Z=1|X)$.

In order to identify the LTE for the subpopulation of compliers, we impose
the following assumptions.

\begin{assumption}
\label{late_a}$(i)$ $\left( Y_{0},Y_{1},T_{1},T_{0},C_{1},C_{0}\right) 
\independent%
Z|X$; $(ii)$ for some $\varepsilon >0$, $\varepsilon \leq e\left( \cdot
\right) \leq 1-\varepsilon ~a.s.,$and $\mathbb{P}\left( T_{1}=1\right) >%
\mathbb{P}\left( T_{0}=1\right) ~a.s.;$ $(iii)$ $\mathbb{P}\left(
T_{1}>T_{0}\right) =1~a.s.;~(iv)$ $\left( Y_{0},Y_{1}\right) $ $%
\independent%
\left( C_{1},C_{0}\right) |T_{1},T_{0},Z$, and for $t\in \left\{ 0,1\right\} 
$, $\mathbb{P}\left( \delta _{t}=1|X,T,Z,Y_{t}\right) =\mathbb{P}\left(
\delta _{t}=1|Y_{t},T,Z\right) ~a.s.$.
\end{assumption}

Assumption \ref{late_a}$\left( i\right) $-$\left( iii\right) $ are standard,
cf. \cite{Abadie2003} and \cite{Frolich2013}\footnote{%
Although standard in the literature, Assumption \ref{late_a}$(iii)$ can be
relaxed, see \cite{DeChaisemartin2014} for details.}. Assumption \ref{late_a}%
$(iv)$ is related to the censoring mechanisms and is the analogous of
Assumption \ref{censoring_identification} in the LTE context; it solves the
additional identification challenge that censoring introduces into the LTE
setup. It is important to notice that Assumption \ref{late_a} does not
restrict how treatment status and instruments affects the censoring
variable, which is weaker than the assumptions commonly used in the
literature, cf. \cite{Frandsen2014}.

In the absence of censoring, \cite{Abadie2003}, \cite{Frolich2007}, and \cite%
{Frolich2013} propose estimators for (\ref{late})-(\ref{lqte}). Although
their procedures are not feasible when $Y$ is censored, we know from the
discussion in Sections \ref{motivating} and \ref{intro-km} that, under
Assumption \ref{late_a}, we can apply the 2SKM procedure to estimate (\ref%
{late})-(\ref{lqte}) in the present context.

The first step towards estimating (\ref{late})-(\ref{lqte}) is to estimate $%
e\left( \cdot \right) $. Noticing that the available instrument $Z$ for $T$
is binary, one can treat $e\left( \cdot \right) $ as an \textquotedblleft
instrumental propensity score\textquotedblright\ and estimate it using
parametric models such as the Logit or Probit specification, or using
nonparametric Kernel or Series estimators as described in the Appendix. We
denote the estimator of $e\left( \cdot \right) $ by $\hat{e}_{n}\left( \cdot
\right) $.

With $\hat{e}_{n}\left( \cdot \right) $ at hands, the next task is to
estimate (\ref{et}) and (\ref{dt}) with censored outcomes. First, there is
no (new) challenge into estimating $\kappa _{t}\left( e\right) $ because one
can simply use its sample analogue,%
\begin{equation*}
\hat{\kappa}_{t,n}\left( \hat{e}_{n}\right) =\frac{1}{n}\sum_{i=1}^{n}\left[ 
\frac{Z_{i}1\left\{ T_{i}=t\right\} }{\hat{e}_{n}\left( X_{i}\right) }-\frac{%
\left( 1-Z_{i}\right) 1\left\{ T_{i}=t\right\} }{1\mathbb{-}\hat{e}%
_{n}\left( X_{i}\right) }\right] .
\end{equation*}%
Next, we plug in $\hat{\kappa}_{t,n}\left( \hat{e}_{n}\right) $ into (\ref%
{et}) and (\ref{dt}), and by using our Kaplan-Meier approach to handle the
censoring problem, we estimate (\ref{et}) and (\ref{dt}) by%
\begin{align}
\mathbb{E}_{n}^{km}\left[ Y_{t}^{c}\right] & =\frac{1}{\hat{\kappa}%
_{t,n}\left( \hat{e}_{n}\right) }\left( \sum_{j=1}^{n_{t1}}W_{jn_{t1}}\frac{%
Q_{j:n_{t1}}}{\hat{e}_{n}\left( X_{\left[ j:n_{t1}\right] }\right) }%
-\sum_{i=1}^{n_{t0}}W_{in_{t0}}\frac{Q_{i:n_{t0}}}{1-\hat{e}_{n}\left( X_{%
\left[ i:n_{t0}\right] }\right) }\right) ,  \label{etn} \\
\hat{F}_{n,Y_{t}^{c}}^{km}\left( y\right) & =\frac{1}{\hat{\kappa}%
_{t,n}\left( \hat{e}_{n}\right) }\left( \sum_{j=1}^{n_{t1}}W_{jn_{t1}}\frac{%
1\left\{ Q_{j:n_{t1}}\leq y\right\} }{\hat{e}_{n}\left( X_{\left[ j:n_{t1}%
\right] }\right) }-\sum_{i=1}^{n_{t0}}W_{in_{t0}}\frac{1\left\{
Q_{i:n_{t0}}\leq y\right\} }{1-\hat{e}_{n}\left( X_{\left[ i:n_{t0}\right]
}\right) }\right) ,  \label{dtn}
\end{align}%
where $n_{tz}=\sum_{i=1}^{n}1\left\{ T=t\right\} 1\left\{ Z=z\right\} $, $%
z\in \left\{ 0,1\right\} $, and for $1\leq i\leq n_{tz},$ $Q_{1:n_{tz}}\leq $
$\cdots \leq Q_{n_{td}:n_{tz}}$ are the ordered $Q$-values in the subsample
with $\left\{ T=t,Z=z\right\} $, $X_{\left[ i:n_{tz}\right] }$ and $\delta _{%
\left[ i:n_{tz}\right] }$ are the $X$ and $\delta $ paired with $Q_{i:n_{tz}}
$, and%
\begin{equation*}
W_{in_{tz}}=\frac{n_{tz}}{n}\frac{\delta _{\lbrack i:n_{tz}]}}{n_{tz}-i+1}%
\prod_{j=1}^{i-1}\left[ \frac{n_{tz}-j}{n_{tz}-j+1}\right] ^{\delta _{\left[
j:n_{tz}\right] }}
\end{equation*}%
is the Kaplan-Meier weights for the subsample with $\left\{ T=t,Z=z\right\} $%
. Once such measures are available, our 2SKM estimators for (\ref{late})-(%
\ref{lqte}) are given by%
\begin{align}
LATE_{n}^{km}& =\mathbb{E}_{n}^{km}\left[ Y_{1}^{c}\right] -\mathbb{E}%
_{n}^{km}\left[ Y_{0}^{c}\right] ,  \label{kmlate} \\
LDTE_{n}^{km}\left( y\right) & =\hat{F}_{n,Y_{1}^{c}}^{km}\left( y\right) -%
\hat{F}_{n,Y_{0}^{c}}^{km}\left( y\right) ,  \label{kmldte} \\
LQTE_{n}^{km}\left( \tau \right) & =\hat{F}_{n,Y_{1}^{c}}^{km,-1}\left( \tau
\right) -\hat{F}_{n,Y_{0}^{c}}^{km,-1}\left( \tau \right) ,  \label{kmlqte}
\end{align}%
where, for $t\in \left\{ 0,1\right\} $, $\mathbb{E}_{n}^{km}\left[ Y_{t}^{c}%
\right] $ is given by (\ref{etn}), $\hat{F}_{n,Y_{t}^{c}}^{km}\left(
y\right) $ is given by (\ref{dtn}) and $\hat{F}_{n,Y_{t}^{c}}^{km,-1}\left(
\tau \right) =\inf \left( y:\hat{F}_{n,Y_{t}^{c}}^{km,r}\left( y\right) \geq
\tau \right) $, where $\hat{F}_{n,Y_{t}^{c}}^{km,r}\left( y\right) $ denotes
the rearrangement of $\hat{F}_{n,Y_{t}^{c}}^{km}\left( y\right) $ if $\hat{F}%
_{n,Y_{t}^{c}}^{km}\left( y\right) $ is not monotone, cf. \cite%
{Chernozhukov2010}\footnote{%
To construct (\ref{kmlate})-(\ref{kmlqte}), we split the sample into four
sub-samples depending on the treatment status $T$ and on the value of the
instrument $D$. This is necessary because Assumptions \ref{late_a}$\left(
iv\right) $-$\left( v\right) $ does not impose any restriction on how $T$
and $D$ affect the censoring probability. If one is willing to strengthen
Assumptions \ref{late_a}$\left( iv\right) $-$\left( v\right) $ to the case
in which these assumptions hold unconditionally on $D$, one would need to
split the sample only on treated and control groups, like in the
unconfoundedness setup. For the sake of generality, we avoid doing so.}.

Next proposition shows that the 2SKM estimators (\ref{kmlate})-(\ref{kmldte}%
) are uniformly consistent, converge weakly, and one can use the bootstrap
to perform asymptotically valid inference. Let 
\begin{eqnarray*}
\tau _{H}^{c} &\equiv &\min_{t}\left\{ \inf \left\{ y:\mathbb{P}\left(
Q_{t}\leq y|T_{1}>T_{0}\right) =1\right\} \right\} , \\
\bar{\tau}^{c} &\equiv &\min_{t}\left\{ \mathbb{P}\left( Y_{t}\leq \tau
_{H}^{c}|T_{1}>T_{0}\right) \right\} .
\end{eqnarray*}

\begin{proposition}
\label{col2}Let Assumption \ref{late_a} be satisfied. Assume that $e\left(
X\right) =$ $\mathbb{P}(Z=1|X)$ satisfy the analogous of either Assumption %
\ref{parametric}, Assumption \ref{kernel}, or Assumption \ref{series} in the
Appendix \ref{pscore-cond}. Then:

$(i)$ If $Y_{1}^{c}$ and $Y_{0}^{c}$ are square integrable, Condition \ref%
{cond1} holds for the subpopulation of compliers, and Assumptions \ref%
{integrability2}-\ref{bias} are satisfied for $\varphi _{z,h_{0}}\left(
\cdot ,\cdot ,\cdot \right) =yde\left( x\right) ^{-1}-y\left( 1-d\right)
\left( 1-e\left( x\right) \right) ^{-1}$, the $ATE_{c}$ estimator (\ref%
{kmlate}) for (\ref{late}) is $\sqrt{n}$-consistent and asymptotically
normal, and the bootstrap procedure described in Theorem \ref{boot} is
asymptotically valid.

$(ii)$ The $DTE_{c}$ $\left( \cdot \right) $ estimator (\ref{ldte}) for (\ref%
{kmldte}) converges weakly in $l^{\infty }\left( \mathcal{Y}\cap (-\infty
,\tau _{H}^{c})\right) ,$ and the bootstrap procedure described in Theorem %
\ref{boot} is asymptotically valid.

$(iii)$ If for the subpopulation of compliers $Y_{0}$ and $Y_{1}$ are
absolutely continuous random variables such that, for $t\in \left\{
0,1\right\} ,$%
\begin{equation*}
0<\inf_{\tau \in \left( 0,\bar{\tau}^{c}\right) }f_{Y_{t}^{c}}\left(
F_{Y_{t}^{c}}^{-1}\left( \tau \right) \right) <\sup_{\tau \in \left( 0,\bar{%
\tau}^{c}\right) }f_{Y_{t}^{c}}\left( F_{Y_{t}^{c}}^{-1}\left( \tau \right)
\right) <\infty ,
\end{equation*}%
the $QTE_{c}$ $\left( \cdot \right) $ estimator (\ref{kmlqte}) for (\ref%
{lqte}) converges weakly in $l^{\infty }(\left( 0,\bar{\tau}^{c})\right) $,
and the bootstrap procedure described in Theorem \ref{boot} is
asymptotically valid.
\end{proposition}

The results in Proposition \ref{col2} are novel to the literature. To the
best of our knowledge, the only related results to Proposition \ref{col2} is 
\cite{Frandsen2014}, who proposes estimators for the distributional and
quantile treatment effects (\ref{ldte}) and (\ref{lqte}), but in the much
simpler setup than ours: \cite{Frandsen2014}'s proposal cannot accommodate
covariates, relies on the censoring variable being always observed, and
requires appropriate support restrictions that excludes from the analysis
some functionals of interest such as the $ATE$. Furthermore, even when \cite%
{Frandsen2014} putative conditions are satisfied, one can show that our 2SKM
estimators are more efficient than his, even though the 2SKM estimator does
not use the full sample of $C_{i}$ values, cf. \cite{Portnoy2010}. These
features highlights the flexibility and power of our proposal.

\subsection{Differences-in-Differences\label{CiC-sec}}

In this section we propose 2SKM estimators for (\ref{att})-(\ref{qtt}) in
the Changes-in-Changes (CIC) setup described in Example 2.3. We make the
following assumptions.

\begin{assumption}
\label{cic-ass}$(i)$ $Y_{0}=h\left( U,I\right) $; $\left( ii\right) $ The
function $h\left( u,j\right) $, where $h:\mathbb{U}\times \left\{
0,1\right\} \rightarrow \mathbb{R},$ is strictly increasing in $u$ for $%
j=\left\{ 0,1\right\} ;$ $\left( iii\right) $ $U%
\independent%
I|G;$ $\left( iv\right) $ The support of $U$ given $G=1$ is contained in the
support of $U$ given $G=0;$ $\left( v\right) $ $\mathbb{P}\left(
G=g,I=j\right) >0$ for all $g,t\in \left\{ 0,1\right\} $; $(vi)$ The four
random variables $Y_{gt}$ have compact support, are absolutely continuous
with densities $f_{Y|G=g,I=j}\left( y\right) $ that are continuously
differentiable and $\inf f_{Y|G=g,I=j}\left( \cdot \right) >0$ and $\sup
f_{Y|G=g,I=j}\left( \cdot \right) <M<\infty ;$ $~(vii)$ $Y$ $%
\independent%
C|I,G$.
\end{assumption}

Assumptions \ref{cic-ass}$(i)$-$\left( vi\right) $ define the CIC classical
setup of \cite{Athey2006}. Assumption \ref{cic-ass}$(vii)$ is related to the
censoring mechanism, and states that conditionally on the group status and
on the time period, the outcome of interest is independent of the censoring
random variable.

The first step towards estimating (\ref{att})-(\ref{qtt}) is to estimate the
nuisance functions $F_{Y_{01}}\left( \cdot \right) $ and $%
F_{Y_{00}}^{-1}\left( \cdot \right) $. Notice that, in contrast with the
unconfoundedness and local treatment effect setups, here the nuisance
functions are affect by the censoring problem. Nonetheless, they can be
estimated by their Kaplan-Meier analogues%
\begin{eqnarray*}
\hat{F}_{n,~Y_{gj}}^{km}\left( y\right)  &=&\sum_{i=1}^{n_{gj}}\tilde{W}%
_{in_{gj}}1\left\{ Q_{i:n_{gj}}\leq y\right\} , \\
\hat{F}_{n,Y_{gj}}^{km,-1}\left( \tau \right)  &=&\inf \left\{ y:\hat{F}%
_{n,~Y_{gj}}^{km}\left( y\right) \geq \tau \right\} ,
\end{eqnarray*}
$g\in \left\{ 0,1\right\} $, $j\in \left\{ 0,1\right\} $, where $n_{gj}$ $%
=\sum_{i=1}^{n}1\left\{ G_{i}=g\right\} 1\left\{ I_{i}=j\right\} ,$ $%
Q_{1:n_{gj}}\leq $ $\cdots \leq Q_{n_{gj}:n_{gj}}$ are the ordered $Q$%
-values in the subsample with $\left\{ G=g,I=j\right\} $, $X_{\left[ i:n_{gj}%
\right] }$ and $\delta _{\left[ i:n_{gj}\right] }$ are the $Q_{i:n_{gj}}$
concomitants, and for $1\leq i\leq n_{gj}$, 
\begin{equation*}
\tilde{W}_{in_{gj}}=\frac{\delta _{\lbrack i:n_{gj}]}}{n_{gj}-i+1}%
\prod_{k=1}^{i-1}\left[ \frac{n_{gj}-k}{n_{gj}-k+1}\right] ^{\delta _{\left[
k:n_{gj}\right] }}.
\end{equation*}%
is the size of the Kaplan-Meier jump for observation $i$ in the subsample
with $\left\{ G=g,I=j\right\} $. Notice that these nonparametric estimators
are fully data-driven, and do not require the use of tuning parameters such
as bandwidths.

With the first-step estimators at hands, we can use our 2SKM approach to
estimate (\ref{att})-(\ref{qtt}). More precisely, we propose to estimate (%
\ref{att})-(\ref{qtt}) by%
\begin{eqnarray}
ATT_{n}^{km} &=&\sum_{i=1}^{n_{11}}\tilde{W}_{in_{11}}Q_{i:n_{11}}-%
\sum_{j=1}^{n_{10}}\tilde{W}_{jn_{10}}\left[ \hat{F}_{n,Y_{01}}^{km,-1}(\hat{%
F}_{n,Y_{00}}^{km}\left( Q_{j:n_{10}}\right) \right] ,  \label{attkm} \\
DTT_{n}^{km}\left( y\right) &=&\hat{F}_{n,Y_{11}}^{km}\left( y\right) -\hat{F%
}_{n,Y_{0}|T=1}^{km}\left( y\right) ,  \label{dttkm} \\
QTT_{n}^{km}\left( \tau \right) &=&\hat{F}_{n,Y_{11}}^{km,-1}\left( \tau
\right) -\hat{F}_{n,Y_{0}|T=1}^{km,r,-1}\left( \tau \right) ,  \label{qttkm}
\end{eqnarray}%
where%
\begin{eqnarray*}
\hat{F}_{n,Y_{0}|T=1}^{km}\left( y\right) &=&\sum_{i=1}^{n_{10}}\tilde{W}%
_{in_{10}}1\left\{ Q_{j:n_{10}}\leq \hat{F}_{n,Y_{00}}^{km,-1}(\hat{F}%
_{n,Y_{01}}^{km}\left( y\right) \right\} , \\
\hat{F}_{n,Y_{0}|T=1}^{km,r,-1}\left( \tau \right) &=&\inf \left( y:\hat{F}%
_{n,Y_{0}|T=1}^{km,r}\left( y\right) \geq \tau \right) ,
\end{eqnarray*}%
and $\hat{F}_{n,Y_{0}|T=1}^{km,r}\left( y\right) $ denotes the rearrangement
of $\hat{F}_{n,Y_{0}|T=1}^{km}\left( y\right) $ if $\hat{F}%
_{n,Y_{0}|T=1}^{km}\left( y\right) $ is not monotone, cf. \cite%
{Chernozhukov2010}.

In the next proposition we show that the 2SKM estimators (\ref{attkm})-(\ref%
{qttkm}) are uniformly consistent, converge weakly, and one can use the
ordinary nonparametric bootstrap to perform asymptotically valid inference.
Let 
\begin{eqnarray*}
\tau _{H}^{cic} &\equiv &\min_{g,j}\left\{ \inf \left\{ y:\mathbb{P}\left(
Q\leq y|G=g,I=j\right) =1\right\} \right\} , \\
\bar{\tau}^{cic} &\equiv &\min_{t}\left\{ \mathbb{P}\left( Y_{t}\leq \tau
_{H}^{cic}\right) \right\} .
\end{eqnarray*}

\begin{proposition}
\label{col3}Let Assumption \ref{cic-ass} be satisfied. Then:

$(i)$ If $Y_{1}$ and $Y_{0}$ are square integrable, Condition \ref{cond1}
holds and Assumptions \ref{integrability2}-\ref{bias} are satisfied for $%
\varphi _{z,h_{0}}\left( \cdot ,\cdot ,\cdot \right) =g(1-j)\left(
y-F_{Y_{01}}^{-1}(F_{Y_{00}}\left( y\right) )\right) $, the $ATT$ estimator (%
\ref{attkm}) for (\ref{att}) is $\sqrt{n}$-consistent and asymptotically
normal, and the bootstrap procedure described in Theorem \ref{boot} is
asymptotically valid.

$(ii)$ The $DTT\left( \cdot \right) $ estimator (\ref{dttkm}) for (\ref{dtt}%
) converges weakly in $l^{\infty }\left( \mathcal{Y}\cap (-\infty ,\tau
_{H}^{cic})\right) ,$ and the bootstrap procedure described in Theorem \ref%
{boot} is asymptotically valid.

$(iii)$ The $QTT\left( \cdot \right) $ estimator (\ref{qttkm}) for (\ref{qtt}%
) converges weakly in $l^{\infty }(\left( 0,\bar{\tau}^{cic})\right) $, and
the bootstrap procedure described in Theorem \ref{boot} is asymptotically
valid.
\end{proposition}

These results in Proposition \ref{col3} are new even when censoring is not
an issue. First, it generalizes \cite{Athey2006} pointwise results to hold
uniformly. Second, it proves that one can use the bootstrap to perform
inference in the CIC setup. Both of these points are of practical relevance: 
$(i)$ because our results hold uniformly, one can test for first-or
second-order stochastic dominance in the same spirit of \cite{Abadie2002}; $%
\left( ii\right) $ by using bootstrapped confidence intervals to conduct
inference on $QTT\left( \cdot \right) $, one completely avoids the need of
estimating density functions to construct standard errors, a task that would
involve choosing tuning parameters. Proposition \ref{col3} shows that these
desirable features naturally carry out to the randomly censored CIC setup.

\section{ Monte Carlo \label{mc}}

In this section, we conduct a small scale Monte Carlo exercise in order to
study the finite sample properties of our proposed policy evaluation
estimators. More precisely, we compare the performance of the two-step
Kaplan-Meier (2SKM) estimators proposed here with those based on $(a)$ the
\textquotedblleft naive\textquotedblright\ approach that uses inverse
probability weighted (IPW) estimators ignoring that the outcome of
interested is subjected to censoring (we label such an approach as
\textquotedblleft Ignore \textquotedblright ); $\left( b\right) $ the
\textquotedblleft naive\textquotedblright\ approach that uses IPW estimators
after dropping all censored data (we label such an approach as
\textquotedblleft Uncens \textquotedblright ); $\left( c\right) $ the \cite%
{Cox1972, Cox1975} Proportional hazard model for the treated and control
groups (we label such an approach as \textquotedblleft Cox\textquotedblright
), in which we exploit the relationship between the conditional hazard
rates, the conditional CDF's, and then integrate out the covariate vector to
get the unconditional CDF's; and $\left( d\right) $ the \cite{Frandsen2014}%
's proposal (we label such an approach as \textquotedblleft
Frandsen\textquotedblright )\footnote{%
For detailed description on how to compute the policy evaluation parameters
using these competing methods, see the Supplemental Appendix.}.
For conciseness, we focus on the unconfoundedness setup.

We consider the following four designs:%
\begin{eqnarray*}
1.~~~Y_{0} &=&e_{0},~Y_{1}=Y_{0}+1, \\
~~C_{0} &=&C_{1}=\varepsilon _{c}; \\
2.~~~Y_{0} &=&e_{0},~Y_{1}=Y_{0}+1+e_{1}, \\
~~C_{0} &=&C_{1}=\varepsilon _{c}; \\
3.~~~Y_{0} &=&X+e_{0},~Y_{1}=Y_{0}+1+X, \\
~~C_{0} &=&C_{1}=\varepsilon _{c};
\end{eqnarray*}%
\begin{eqnarray*}
4.~~~Y_{0} &=&X+e_{0},~Y_{1}=Y_{0}+1+X+e_{1}, \\
~~C_{0} &=&C_{1}=\varepsilon _{c},
\end{eqnarray*}%
where $X,\varepsilon _{0}$ and $\varepsilon _{1}$ are independently
distributed as standard normals, and $\varepsilon _{c}$ is independently
distributed as exponential with parameter $a_{c}$,\ where $a_{c}$ is chosen
such that the percentage of censoring in the sample is approximately equal
to 10 or 30 percent. Note that, because $\varepsilon _{c}$ is exponentially
distributed, censoring is more concentrated on the upper tail of the
distribution, as is typically the case. All designs are adapted from \cite%
{Frandsen2014}. Design $1$ is the baseline setup, in which potential
outcomes do not depend on covariates, and the treatment effect is homogenous
(constant) across the entire distribution. Design $2$ introduces
heterogeneity by allowing the policy intervention to affect both the mean
and the variance of the potential outcomes, whereas Design $3$ introduces
heterogeneity by allowing potential outcomes to depend on covariates $X$.
Design $4$ is the most \textquotedblleft heterogeneous\textquotedblright\
design: it combines Designs $2$ and $3$. In all designs, $P\left(
T=1|X\right) =\exp (0.5X)/(1+\exp \left( 0.5X\right) )$, and $\mathbb{E}%
\left( Y_{1}\right) =F_{Y_{1}}^{-1}\left( 0.5\right) =1$, $\mathbb{E}\left(
Y_{0}\right) =F_{Y_{0}}^{-1}\left( 0.5\right) =0$, implying that the $%
ATE=QTE\left( 0.5\right) =1$. The observed data is $\left\{ Q_{i},\delta
_{i},X_{i},T_{i}\right\} _{i=1}^{n}$, where $Q_{i}=\min \left(
Y_{i},C_{i}\right) $ and $\delta _{i}=1\left\{ Y_{i}\leq C_{i}\right\} $.
Nonetheless, in order to use \cite{Frandsen2014} approach, we assume that $%
C_{i}$ is observed for both censored and uncensored observations, though we
do not need such restrictive condition to compute the 2SKM, the
\textquotedblleft naive approaches\textquotedblright , or the Cox based
estimators.

The finite sample comparisons are based on bias\footnote{%
In the Supplemental Appendix  we also compare the root mean
square errors of the competing methods.} for $\mathbb{E}\left( Y_{1}\right) $%
, $\mathbb{E}\left( Y_{0}\right) ,$ $F_{Y_{1}}^{-1}\left( 0.5\right) $, $%
F_{Y_{0}}^{-1}\left( 0.5\right) ,$ $ATE$ and $QTE\left( 0.5\right) $. When
censoring is not present, the 2SKM estimators are numerically equivalent to
those base on the \textquotedblleft naive approaches\textquotedblright .
Thus, we report only the 2SKM, Cox, and \cite{Frandsen2014} estimators in
these simulation setups. All simulations are based on a thousand Monte Carlo
experiments, with a sample size of $n=1,000$ across all scenarios. We
estimate $p\left( \cdot \right) $ using \cite{Hirano2003} series logit
estimator with $1,X,X^{2},X^{3}$ as power functions.

The simulation results are presented in Table \ref{tab:bias}. The
simulations show that the proposed 2SKM estimators for $\mathbb{E}\left(
Y_{1}\right) ,$ $\mathbb{E}\left( Y_{0}\right) ,$ $F_{Y_{1}}^{-1}\left(
0.5\right) $, $F_{Y_{0}}^{-1}\left( 0.5\right) ,$ $ATE$ and $QTE\left(
0.5\right) $ have minimal bias across all DGP's, and outperforms all other
methods, specially when covariates play an important role. This is not
surprising, since the 2SKM approach is the only appropriate method to
estimate all measures of interest in the presence of censoring and
covariates, without relying on functional form assumptions. Even when the
potential outcomes do not depend on covariates, however, our proposed 2SKM
estimators perform nearly as well as \cite{Frandsen2014}'s estimators, even
though we make use of less information (we do not use $C_{i}^{\prime }s$
whatsoever). Such a feature stress the flexibility and appeal of our 2SKM
estimators.

\begin{table}[!ht]
\caption{Simulated bias, in percentage points, under the unconfoundedness
setup}
\label{tab:bias}%
\begin{adjustbox}{width=1.\textwidth}
\par
\begin{threeparttable}
    \begin{tabular}{crrrrrrrrrrrrrrr}
    \toprule \cmidrule{1-16}
    \multicolumn{16}{c}{DGP=1} \\
    \midrule
          & \multicolumn{3}{c}{Not Censored} &       & \multicolumn{5}{c}{Censoring=10\%}    &       & \multicolumn{5}{c}{Censoring=30\%} \\ \cmidrule{2-4} \cmidrule{6-10}  \cmidrule{12-16} 
    Objects / Estimators & 2SKM  & Cox   & Frandsen &       & 2SKM  & Ignore & Uncens & Cox   & Frandsen &       & 2SKM  & Ignore & Uncens & Cox   & Frandsen \\ \cmidrule{2-4} \cmidrule{6-10}  \cmidrule{12-16}
     $\mathbb{E}\left( Y_{1}\right)$  & 0.08  & 0.36  & 0.07  &       & -0.07 & -13.03 & -16.61 & 0.41  & -0.11 &       & -1.37 & -42.03 & -56.50 & 0.74  & -0.37 \\
     $\mathbb{E}\left( Y_{0}\right)$  & 0.09  & 0.42  & 0.10  &       & -0.32 & -3.71 & -7.62 & 0.07  & -0.29 &       & -0.24 & -11.72 & -29.50 & 0.66  & -0.02 \\
     $F_{Y_{1}}^{-1}\left( 0.5\right)$ & -0.14 & 0.06  & -0.09 &       & -0.07 & -16.76 & -13.10 & -32.91 & 0.20  &       & -0.63 & -50.03 & -50.26 & -86.41 & 0.35 \\
     $F_{Y_{0}}^{-1}\left( 0.5\right)$ & -0.08 & 0.16  & 0.06  &       & -0.17 & -0.47 & -12.34 & -0.64 & -0.01 &       & -0.40 & -1.00 & -38.98 & -2.17 & 0.14 \\
    $ATE$    & 0.00  & -0.06 & -0.03 &       & 0.25  & -9.32 & -9.00 & 0.34  & 0.18  &       & -1.13 & -30.31 & -27.01 & 0.08  & -0.35 \\
    $QTE(0.5)$    & -0.06 & -0.09 & -0.15 &       & 0.10  & -16.29 & -0.76 & -32.27 & 0.21  &       & -0.24 & -49.03 & -11.28 & -84.24 & 0.21 \\ \cmidrule{1-16}
    \multicolumn{16}{c}{DGP=2} \\ \cmidrule{1-16}
          & \multicolumn{3}{c}{Not Censored} &       & \multicolumn{5}{c}{Censoring=10\%}    &       & \multicolumn{5}{c}{Censoring=30\%} \\ \cmidrule{2-4} \cmidrule{6-10}  \cmidrule{12-16}
    Objects / Estimators & 2SKM  & Cox   & Frandsen &       & 2SKM  & Ignore & Uncens & Cox   & Frandsen &       & 2SKM  & Ignore & Uncens & Cox   & Frandsen \\ \cmidrule{2-4} \cmidrule{6-10}  \cmidrule{12-16}
     $\mathbb{E}\left( Y_{1}\right)$  & -0.19 & 0.27  & -0.18 &       & -0.13 & -17.86 & -25.74 & 0.57  & -0.14 &       & -2.81 & -54.10 & -79.80 & 0.77  & -1.16 \\
     $\mathbb{E}\left( Y_{0}\right)$  & -0.17 & 0.12  & -0.19 &       & -0.15 & -3.58 & -7.57 & 0.20  & -0.15 &       & -0.55 & -11.97 & -29.87 & 0.53  & -0.15 \\
     $F_{Y_{1}}^{-1}\left( 0.5\right)$ & -0.69 & -0.41 & -0.63 &       & -0.53 & -22.19 & -23.23 & -42.83 & -0.05 &       & -2.05 & -57.88 & -78.66 & -97.45 & -0.26 \\
     $F_{Y_{0}}^{-1}\left( 0.5\right)$ & -0.41 & -0.25 & -0.41 &       & -0.55 & -0.81 & -13.25 & -0.91 & -0.36 &       & -0.74 & -1.20 & -39.15 & -2.33 & -0.12 \\
    $ATE$    & -0.01 & 0.15  & 0.01  &       & 0.02  & -14.29 & -18.16 & 0.37  & 0.01  &       & -2.26 & -42.13 & -49.93 & 0.25  & -1.01 \\
    $QTE(0.5)$    & -0.28 & -0.17 & -0.22 &       & 0.02  & -21.38 & -9.98 & -41.92 & 0.30  &       & -1.31 & -56.69 & -39.50 & -95.12 & -0.14 \\ \cmidrule{1-16}
    \multicolumn{16}{c}{DGP=3} \\ \cmidrule{1-16}
          & \multicolumn{3}{c}{Not Censored} &       & \multicolumn{5}{c}{Censoring=10\%}    &       & \multicolumn{5}{c}{Censoring=30\%} \\ \cmidrule{2-4} \cmidrule{6-10}  \cmidrule{12-16}
    Objects / Estimators & 2SKM  & Cox   & Frandsen &       & 2SKM  & Ignore & Uncens & Cox   & Frandsen &       & 2SKM  & Ignore & Uncens & Cox   & Frandsen \\ \cmidrule{2-4} \cmidrule{6-10}  \cmidrule{12-16}
     $\mathbb{E}\left( Y_{1}\right)$  & 0.12  & -2.47 &47.518  &       & -0.28 & -23.16 & -36.44 & -1.30 & 46.85 &       & -4.58 & -69.79 & -113.51 & 2.31  & 45.16 \\
     $\mathbb{E}\left( Y_{0}\right)$  & -0.06  & -4.33 & 23.68 &       & -0.03 & -5.04 & -10.87 & -3.64 & -23.64 &       & -0.89 & -17.73 & -43.95 & -2.14 & -23.70 \\
     $F_{Y_{1}}^{-1}\left( 0.5\right)$ & 0.00  & 102.16  & 47.46  &       & -0.77 & -24.89 & -36.74 & 28.84 & 47.06 &       & -2.40 & -60.71 & -112.99 & -77.31 & 47.96 \\
     $F_{Y_{0}}^{-1}\left( 0.5\right)$ & -0.14 & -38.78 & -23.74 &       & -0.22 & -0.55 & -18.46 & -39.38 & -23.59 &       & -0.89 & -1.35 & -55.52 & -41.93 & -23.60 \\
    $ATE$    & 0.18  & 1.86  & 71.19  &       & -0.25 & -18.11 & -25.58 & 2.35  & 70.48 &       & -3.69 & -52.06 & -69.57 & 4.45  & 68.86 \\
    $QTE(0.5)$    & 0.014  & 140.94  & 71.2  &       & -0.55 & -24.34 & -18.28 & 68.22 & 70.65 &       & -1.51 & -59.36 & -57.46 & -35.38 & 71.56 \\ \cmidrule{1-16}
    \multicolumn{16}{c}{DGP=4} \\ \cmidrule{1-16}
          & \multicolumn{3}{c}{Not Censored} &       & \multicolumn{5}{c}{Censoring=10\%}    &       & \multicolumn{5}{c}{Censoring=30\%} \\ \cmidrule{2-4} \cmidrule{6-10}  \cmidrule{12-16}
    Objects / Estimators & 2SKM  & Cox   & Frandsen &       & 2SKM  & Ignore & Uncens & Cox   & Frandsen &       & 2SKM  & Ignore & Uncens & Cox   & Frandsen \\ \cmidrule{2-4} \cmidrule{6-10}  \cmidrule{12-16}
     $\mathbb{E}\left( Y_{1}\right)$  & -0.20 & -1.81 & 46.85 &       & 0.05  & -25.77 & -41.60 & 0.36  & 47.30 &       & -6.88 & -77.34 & -127.54 & 3.37  & 43.98 \\
     $\mathbb{E}\left( Y_{0}\right)$  & 0.00  & -4.29 & -23.51 &       & -0.16 & -5.11 & -11.04 & -3.76 & -23.78 &       & -1.34 & -18.00 & -44.39 & -2.23 & -24.11 \\
     $F_{Y_{1}}^{-1}\left( 0.5\right)$ & -0.95 & 102.73 & 46.33 &       & -0.72 & -26.43 & -42.17 & 25.08 & 46.94 &       & -4.29 & -63.28 & -126.18 & -81.33 & 46.79 \\
     $F_{Y_{0}}^{-1}\left( 0.5\right)$ & -0.26 & -38.65 & -23.74 &       & -0.39 & -0.68 & -19.11 & -39.46 & -23.86 &       & -1.10 & -1.37 & -55.42 & -41.72 & -23.53 \\
    $ATE$    & -0.21 & 2.49  & 70.36 &       & 0.21  & -20.66 & -30.56 & 4.12  & 71.08 &       & -5.54 & -59.33 & -83.15 & 5.59  & 68.09 \\
    $QTE(0.5)$    & -0.69 & 141.38 & 70.07 &       & -0.33 & -25.75 & -23.06 & 64.54 & 70.80 &       & -3.19 & -61.91 & -70.76 & -39.60 & 70.32 \\ 
    \bottomrule
    \end{tabular}%
\par
\begin{tablenotes}[para,flushleft]
\small{
Note: Simulations based on one thousand Monte Carlo experiments. Sample size equal to 1,000. ``2SKM'' stands for estimators based on our proposal. ``Ignore'' stands for estimators based on inverse probability weight (IPW) estimators that ignore the censoring problem. ``Uncens'' stands for IPW estimators after dropping all censored outcomes. Cox stands for estimators based on 
the Cox-Proportional hazard model for the treated d and control groups. ``Frandsen'' stands for the estimators based on \cite{Frandsen2014}. }
\end{tablenotes}
\end{threeparttable}
\end{adjustbox}
\end{table}

As expected, the \textquotedblleft naive\textquotedblright\ estimators that
ignore the censoring issue, or use only uncensored observations are severely
biased. Another feature worth mentioning is that estimators for $\mathbb{E}%
\left( Y_{1}\right) $, $\mathbb{E}\left( Y_{0}\right) $ and $ATE$ based on
the Cox proportional hazard model have close to minimal bias, even though
the model is misspecified (the conditional hazards are not proportional in
the analyzed DGP's). However, the same is not true for the Cox estimators
for the median. As discussed by \cite{Portnoy2003}, this is due to the fact
that the Cox model greatly restricts the behavior of the quantile effects,
leading to inconsistent and severely biased estimates when the underlying
assumptions of the model are not satisfied, as it is the case here. Finally,
notice that \cite{Frandsen2014}'s estimators are unbiased in DGP's $1$ and $2
$, but are severely biased in DGP's $3$ and $4$. This is a simple
consequence of \cite{Frandsen2014} not being able to accommodate covariates
into the analysis, which turns out to be crucial in the last two DGP's.

In summary, our simulations highlights that our proposed 2SKM estimators
exhibit very good finite sample properties in all analyzed designs. On the
other hand, ignoring the censoring problem, imposing ad hoc functional form
restrictions in the distribution of the potential outcomes, or not
accommodating covariates into the analysis may lead to spurious conclusions
about the policy effectiveness.

\section{Conclusion\label{conc}}

In this paper we proposed a class of Kaplan-Meier two-step estimators when
the outcome of interest is subjected to right-censoring mechanisms. We
provided sufficient conditions for the 2SKM estimator to be uniformly
consistent and converge weakly to a tight Gaussian process with mean zero,
and covariance function that may depend on the underlying DGP in rather
complicated ways. To conduct asymptotically valid inference, we have shown
that one can use the ordinary nonparametric bootstrap. We illustrate the
relevance and applicability of our general results by proposing new average,
distributional, and quantile treatment effects estimators under the
unconfoundedness, local treatment effect, and changes-in-changes setups with
censored outcomes.

Although we have focused on the aforementioned three policy setups, the
proposed 2SKM tools can be applied to other designs such as multi-valued
treatments, cf. \cite{Cattaneo2010}; dynamic treatment effects, cf. \cite%
{Sianesi2004}, \cite{Fredriksson2008}, \cite{vandenberg2009}, and \cite%
{Vikstrom2014}; fuzzy differences-in-differences, cf. \cite%
{DeChaisemartin2015}; distributional differences-in-differences, cf. \cite%
{Callaway2015}, and \cite{Callaway2015a}; and also to identify other
parameters of interest such as the marginal treatment effects, cf. \cite%
{Heckman2001, Heckman2005}; or to conduct Oaxaca-Blinder-type
decompositions, cf. \cite{Fortin2011} for a review, and \cite%
{Garcia-Suaza2015} for related results with censored outcomes. In short, in
this paper we have shown that, by using the 2SKM approach, many policy
evaluation tools available for \textquotedblleft complete
data\textquotedblright\ can be extended to accommodate randomly censored
outcomes.

\appendix
\renewcommand{\theassumption}{A.\arabic{assumption}}

\section*{Appendix: Conditions on the propensity score\label{pscore-cond}}

{\footnotesize {\ In this appendix, we discuss sufficient conditions for the
propensity score such that the 2SKM estimators presented in Sections \ref%
{unc-ex} and \ref{lte-ex} converge weakly to a Gaussian process and the
ordinary nonparametric bootstrap is valid. We discuss three alternative
approaches to estimate the propensity score $p\left( \cdot \right) $: the
parametric approach (e.g. Logit or Probit specifications), the nonparametric
kernel-based approach, and the nonparametric series approach. } }

{\footnotesize We start our discussion with the most popular case in applied
research, the parametric approach. If one decided to follow this procedure
and use a flexible parametric estimator of $p\left( \cdot \right) $, the
2SKM estimators discussed in the main text will have the desirable
properties as long as the following condition is met.  }
\footnotesize
\begin{assumption}
{\footnotesize {{\label{parametric} Assume that $p=p\left( \cdot ;\theta
_{0}\right) \in \left( 0,1\right) ,$ where $\theta _{0}\in \Theta \subset 
\mathbb{R}^{d_{\theta }}$ with $d_{\theta }$ being a positive integer, is a
smooth function of $\theta _{0}$ with uniformly continuous, bounded, and
square integrable first derivative, $\dot{p}\left( \cdot ;\theta \right) ,$
with respect to $\theta $. Furthermore there exist a strong consistent
estimator $\hat{\theta}_{n}$ for $\theta _{0}$ that satisfies the following
linear expansion:%
\begin{equation*}
\sqrt{n}\left( \hat{\theta}_{n}-\theta _{0}\right) =\frac{1}{\sqrt{n}}%
\sum_{i=1}^{n}I_{\theta _{0}}\left( T_{i},X_{i}\right) +o_{\mathbb{P}}(1),
\end{equation*}%
where $I_{\theta _{0}}\left( \cdot \right) $ is such that $(i)$ $\mathbb{E}%
\left[ I_{\theta _{0}}\left( T,X\right) \right] =0,$ and $L_{\theta
_{0}}\equiv \mathbb{E}\left[ I_{\theta _{0}}\left( T,X\right) I_{\theta
_{0}}\left( T,X\right) ^{^{\prime }}\right] $ exists and is positive
definite $(ii)$ $\lim_{\varepsilon \rightarrow 0}\mathbb{E}\left[
\sup_{\theta \in \Theta _{0}\left\vert ,\theta -\theta _{0}\right\vert \leq
\varepsilon }\left\vert I_{\theta }\left( T,X\right) -I_{\theta _{0}}\left(
T,X\right) \right\vert \right] =0$ where $\Theta _{0}$ is a neighborhood of $%
\theta _{0}$, $\Theta _{0}\subset \Theta $, and $(iii)$ the class of
functions $\{I_{\theta }\left( \cdot ,\cdot \right) ,\theta \in \Theta \}$
is Donsker. } }  }
\end{assumption}

{\footnotesize \ Condition \ref{parametric} is very mild. It assumes that
the propensity score is smooth, and one can find a $\sqrt{n}$-consistent
asymptotically linear estimator $\theta _{n}$ for $\theta _{0}$. For
instance, these assumptions are satisfied when the $p\left( \cdot ;\theta
_{0}\right) $ has a logit or probit specification.  }

{\footnotesize Although using parametric methods to estimate the propensity
are straight-forward and popular in applied research, one must have in mind
that misspecification of the propensity score may lead to severe distortion
on the policy evaluation parameters of interest, cf. \cite{Huber2013}. To
overcome this potential drawback, one can use nonparametric estimators for
the propensity score. Next we discuss two alternative estimators for the
propensity score: the leave-one-out Nadaraya-Watson kernel-based estimator,
and the Logit Series estimator.  }

{\footnotesize {We first consider the kernel-based estimator. Let $K\left(
\cdot \right) $ be a Kernel function and $b$ be a bandwidth. The
leave-one-out Nadaraya-Watson estimator of the propensity score is given by }
}

{\footnotesize {\ 
\begin{equation*}
\hat{p}_{n}^{\ker }\left( X_{i}\right) =\frac{\frac{1}{nb^{k}}%
\sum_{j=1,i\not=j}^{n}T_{i}K\left( \frac{X_{j}-X_{i}}{b}\right) }{\frac{1}{%
nb^{k}}\sum_{j=1}^{n}K\left( \frac{X_{j}-X_{i}}{b}\right) }.
\end{equation*}%
We impose the following sufficient conditions when one chooses to estimate $%
p $ following this kernel approach. }  }

\begin{assumption}
{\footnotesize {\label{kernel}$(i)$The support of $X$, $\mathcal{X}$, is a
Cartesian product of compact intervals and the density of $X$,\ $f\left(
x\right) $, is bounded away from zero on $\mathcal{X}$; $(ii)$ for all $x\in 
\mathcal{X}$, $p\left( x\right) \in \left( 0,1\right) $ and is continuously
differentiable of order $s>k$; $(iii)$ $K\left( u\right) $ is a kernel of
order $s$, is symmetric around zero, is equal to zero outside $%
\prod\limits_{i=1}^{k}\left[ -1,1\right] $, and is continuously
differentiable; $(iv)$\ as $n\rightarrow \infty $, the bandwidth $%
b\rightarrow 0$, and $\log \left( n\right) /\left( nb\right)
^{k+s}\rightarrow 0$. }  }
\end{assumption}

{\footnotesize {Similar assumptions have been adopted by \cite{Heckman1998b}%
, \cite{Ichimura2005}, among others. Assumption \ref{kernel}$(i)$ restricts
the distribution of $X$ and requires that all covariates are continuous.
Nonetheless, at the expense of additional notation, we can deal with the
case where $X$ has both continuous and discrete components by means of
sample splitting based on the discrete covariates. Assumption \ref{kernel}$%
(ii)$ requires sufficient smoothness of the propensity score, whereas
Assumption \ref{kernel}$(iii)$ and $($$iv)$ impose some conditions on the
kernel and bandwidths. This assumption guarantees the asymptotic
negligibility of higher order terms. } }

{\footnotesize {Now, we move our discussion to series logit estimator (SLE)
introduced by \cite{Hirano2003}. To define the SLE, let $\lambda =\left(
\lambda _{1},\dots ,\lambda _{r}\right) ^{\prime }$ be a $r$-dimensional
vector of \ non-negative integers with norm $\left\vert \lambda \right\vert
=\sum_{j=1}^{r}\lambda _{j}$. Let $\left\{ \lambda \left( l\right) \right\}
_{l=1}^{\infty }$ be a sequence including all distinct multi-indices $%
\lambda $ such that $\left\vert \lambda \left( l\right) \right\vert $ is
non-decreasing in $l$ and let $x^{\lambda }=\prod_{j=1}^{r}x_{j}^{\lambda
_{j}}$. For any integer $L$, define $R^{L}\left( x\right) =\left( x^{\lambda
\left( 1\right) },\dots ,x^{\lambda \left( L\right) }\right) ^{\prime }$ as
a vector of power functions. Let $\mathcal{L}\left( a\right) =\exp \left(
a\right) /\left( 1+\exp \left( a\right) \right) $ be the logistic $CDF$. The
SLE for $p\left( x\right) $ is defined as $\hat{p}_{n}^{\func{series}}\left(
x\right) =\mathcal{L}\left( R^{L}\left( x\right) ^{\prime }\pi _{n,L}\right) 
$, where 
\begin{equation*}
\pi _{n,L}=\arg \max_{\pi _{L}}\frac{1}{n}\sum_{i=1}^{n}D_{i}\log \left( 
\mathcal{L}\left( R^{L}\left( X_{i}\right) ^{\prime }\pi _{L}\right) \right)
+\left( 1-D_{i}\right) \log \left( 1-\mathcal{L}\left( R^{L}\left(
X_{i}\right) ^{\prime }\pi _{L}\right) \right) .
\end{equation*}%
}  }

{\footnotesize {\ Following \cite{Hirano2003}, we impose the following
sufficient conditions when one use the SLE. }  }

\begin{assumption}
{\footnotesize {\label{series}$(i)$ The support of $X$, $\mathcal{X}$, is
Cartesian product of compact intervals and the density of $X$, $f\left(
x\right) $, is bounded away from zero on $\mathcal{X}$; $(ii)$ for all $x\in 
\mathcal{X}$, $p\left( x\right) \in \left( 0,1\right) $ is continuously
differentiable of order $s\geq 13k$; ($iii)$ the series logit estimator of $%
p\left( x\right) $ uses a power series with $L=a\cdot N^{v}$ for some $a>0$
and $1/\left( s/k-2\right) <v<1/11$. }  }
\end{assumption}

{\footnotesize {Similar assumptions have been adopted by \cite{Hahn1998}, 
\cite{Hirano2003}, \cite{Donald2013}, among others. Assumption \ref{series}$%
(i)$ is the same as Assumption \ref{kernel}$(i)$. Assumption \ref{series}$%
(ii)$ requires sufficient smoothness of the propensity score, whereas
Assumption \ref{series}$(iii)$ restricts the rate at which additional terms
are added to the series approximation of $p\left( x\right) $, depending on
the dimension of $X$ and the number of derivatives of $p\left( x\right) $.
The restriction on the derivatives in Assumption \ref{series}$(iii)$
guarantees the existence of a $v$ that satisfy the conditions in Assumption %
\ref{series}$(ii)$. } }

{\footnotesize {When comparing Assumption \ref{series} with Assumption \ref%
{kernel}, one can see that the SLE requires the propensity score to be
smoother than the Nadaraya-Watson kernel estimator. Nonetheless $\hat{p}%
_{n}^{\func{series}}\left( x\right) $ is automatically bounded between zero
and one and the curse of dimensionality affects its practical
implementability less severely than $\hat{p}_{n}^{\ker }\left( \cdot \right) 
$. Therefore, the choice between estimators for the propensity score should
depend on the application one has at hands. }  }

{\footnotesize 
\bibliographystyle{ecta}
\bibliography{JMP}

\end{document}